\documentclass[pra,twocolumn,showpacs]{revtex4-2}

\usepackage{amsmath}
\usepackage{graphicx}
\usepackage{ulem}
\usepackage{dcolumn}
\usepackage{epsfig}
\usepackage{bm}
\usepackage{array}
\usepackage[frenchb]{babel}
\usepackage{hyperref}
\hypersetup{
colorlinks=true,
citecolor=black, 
linkcolor=red,
urlcolor=black,
bookmarks=true,
pdfmenubar=true
}

\usepackage{hyperref}
\usepackage{graphicx}
\usepackage{amsmath}
\usepackage{amsfonts}
\usepackage{amssymb}
\usepackage{xcolor}
\usepackage{bbm}
\usepackage{calrsfs}
\usepackage{dutchcal}
\usepackage{amsthm}

\newcommand{\be}{\begin{equation}}
\newcommand{\ee}{\end{equation}}

\newcommand{\beq}{\begin{eqnarray}}
\newcommand{\eeq}{\end{eqnarray}}

\newcommand{\bra}[1]{\ensuremath{\langle#1|}}
\newcommand{\ket}[1]{\ensuremath{|#1\rangle}}

\newcommand{\mean}[1]{\ensuremath{\big\langle #1 \big\rangle}}

\begin{document}
\title{Multiparticle Entanglement Dynamics of Quantum Chaos \\
in a Bose-Einstein condensate}
\author{Sheng-Chang Li$^{1,2}$}
\author{Luca Pezz\`{e}$^{2}$}
\author{Augusto Smerzi$^{2}$}
\affiliation{1. MOE Key Laboratory for Nonequilibrium Synthesis and Modulation of Condensed Matter, Shaanxi Province Key Laboratory of Quantum Information and Quantum Optoelectronic Devices, and School of Physics, Xi'an Jiaotong University, Xi'an 710049, China\\
2. QSTAR, INO-CNR and LENS, Largo Enrico Fermi 2, Firenze 50125, Italy}

\begin{abstract}
We study the particle-entanglement dynamics witnessed by the quantum Fisher information (QFI) 
of a trapped Bose-Einstein condensate governed by the kicked rotor Hamiltonian. 
The dynamics is investigated with a beyond mean-field approach.
We link the time scales of the validity of this approximation in, both, classical regular and 
chaotic regions, with the maximum Lyapunov exponents of the classical system.
This establishes an effective connection between the classical chaos and the QFI. 
We finally study the critical point of a quantum phase transition using the beyond mean-field approximation by considering
a two-mode bosonic Josephson junction with attractive interparticle interaction.
\end{abstract}

\maketitle

\section{Introduction}
Quantum chaos aims to investigate the properties of quantum systems that have a chaotic dynamics in the 
classical limit. 
First works on quantum chaos can be traced back to 1950s \cite{Wigner1951} or even earlier \cite{Einstein1917}.
While systematic theoretical studies emerged in the late 1980s \cite{Bychek2019}, the experimental activity -- requiring 
highly controllable many-body systems -- has begun in the last years~\cite{ChaudhuryNATURE2009}.
Atomic Bose-Einstein condensates (BECs) can be trapped in time-dependent potentials having almost arbitrary geometries and are therefore 
emerging as ideal candidates for the study of quantum chaos in many-body systems~\cite{Stockmann1990,Graf1992,Milner2001,Friedman2001}. Several efforts have been devoted to 
investigate chaotic dynamics in both scalar and spinor BECs, including driven two-mode \cite{Zhang2008}, tilted triple-well \cite{Lepers2008}, kicked ring  \cite{Liu2006}, kicked optical-lattice \cite{Martin2008}, barrier-height-modulated double-well \cite{Kidd2019}, dissipative double-well \cite{Coullet2002}, dissipative optical-lattice \cite{Bychek2019}, spin-1 \cite{Cheng2010}, spin-2 \cite{Kronjager2008}, and coupled spinor atom-molecule BECs \cite{Cheng2009}. 
A paradigmatic model to study quantum chaos is the kicked rotor~\cite{Haake1987, HaakeBOOK} 
which can be realized with trapped BECs \cite{Moore1995, Fallani2004, Christiani2004, Duffy2004, Creffield2006}. 
This model has been widely studied to illustrate classical and quantum chaos and the transition between them \cite{Chirikov1979,Stockmann1999}.

An intriguing problem emerging in the recent literature is the study of the interplay between 
quantum chaos and entanglement. 
It has been indeed recognized that chaotic dynamics and entanglement generation are 
intimately related in many systems \cite{LerosePRA2020b, GietkaPRB2019, Lewis-Swan2019}.
In particular, Ref.~\cite{LerosePRA2020b} presented an unifying picture connecting 
the multiparticle entanglement growth to the quantifiers of classical and quantum chaos.
In this paper, we study the multiparticle entanglement dynamics of a BEC governed by the 
quantum kicked rotor (QKR) model. 
Multipartite entanglement is witnessed by studying the 
quantum Fisher information (QFI)~\cite{PezzePRL2009,note_multipartite,TothPR2009,HyllusPRA2012, TothPRA2012, TothJPA2014, PezzePNAS2016, PezzeRMP2018}.
The QFI~\cite{HelstromBOOK, CavesPRL1994} has proved a quite valuable quantity for the analysis of entanglement in complex systems~\cite{LerosePRA2020b, GietkaPRB2019, MaPRA2009, SorelliPRA2019, HaukeNATPHY2016, PezzePRL2017, PappalardiJSM2017, GabbrielliSCIREP2018, GabbrielliNJP2019, FrerotNATCOMM2019}. 
Here, we study the chaotic dynamics of the QFI under the framework of a beyond mean-field (BMF) approximation~\cite{AnglinPRA2001b}.
We show that the time scalings of the validity of the BMF dynamics are related with the maximum  Lyapunov exponents of the corresponding classical chaotic regimes. 
This provides an effective relation between the time scales of quantum evolution in a chaotic system and the QFI. 
We further employ the BMF method to the analysis of the ground state of the system.
 We show that the QFI evaluated within the BMF method is in excellent agreement with exact results able to characterize the 
 quantum phase transition in finite-size systems. 
The multipartite entanglement detected by the QFI is a resource for parameter estimation beyond the sensitivity limit achievable with separable (or classically-correlated) states~\cite{PezzePRL2009, HyllusPRA2012, TothPRA2012, TothJPA2014, PezzePNAS2016, PezzeRMP2018}.
In this respect, our results show that the generation of metrologically useful entanglement is much faster in chaotic systems, 
when compared to entanglement generation in integrable systems. 
As shown in Ref.~\cite{Fiderer2018}, quantum chaos can also enhance the sensitivity in the estimation of Hamiltonian parameters.

\section{Quantum Fisher information within a beyond mean-field approach}

In the following, we study the dynamics of a two-mode model governed by the Hamiltonian ($\hbar=1$)
\begin{equation}\label{TMTDH}
\hat{H}(t)=a(t)\hat{J}_x+\frac{c}{N}\hat{J}_z^2,
\end{equation}
where 
$\hat{J}_x= (\hat{a}_1^\dagger\hat{a}_2+\hat{a}_2^\dagger\hat{a}_1)/2$, 
$\hat{J}_y = (\hat{a}_1^\dagger\hat{a}_2-\hat{a}_2^\dagger\hat{a}_1)/2{i}$, and  
$\hat{J}_z = (\hat{a}_1^\dagger\hat{a}_1-\hat{a}_2^\dagger\hat{a}_2)/2$ 
are the SU(2) generators 
satisfying angular momentum commutation relation $[\hat{J}_x,\hat{J}_y]=i\hat{J}_z$ with cyclic permutation.
Here, $\hat{a}_{1,2}$ ($\hat{a}_{1,2}^\dagger$) are bosonic annihilation
(creation) operators for the two modes (labelled as ``1'' and ``2''). 
The time-dependent parameter $a(t)$ describes the coupling between the two modes, $c$ depends on the two-body interaction, and $N$ is the total number of particles.
In atomic BEC experiments, the two-mode model (\ref{TMTDH}) can be realized by using either a double-well trapping potential \cite{Javanainen1986, Milburn1997, Zapata1998, Smerzi1997, JavanainenPRA1999, SchummNATPHYS2005, Pezze2005, AlbiezPRL2005, Ananikian2006, Gati2007, TrankwalderNATPHYS2016, SpagnolliPRL2017} or two hyperfine atomic states \cite{Cirac1998, Steel1998, ZiboldPRL2010, Gross2010}. As initial state, we consider a coherent spin state~\cite{ArecchiPRA1972} 
\beq
\ket{\psi_s} =&&  \sum_{\mu=-\frac{N}{2}}^{\frac{N}{2}} \sqrt{\bigg(\begin{array}{c}
    N  \\
   \frac{N}{2}+\mu   
\end{array}\bigg)} \bigg( \cos \frac{\theta}{2} \bigg)^{\frac{N}{2}-\mu} \bigg( \sin \frac{\theta}{2} \bigg)^{\frac{N}{2}+\mu} \nonumber \\
&& \times e^{i (\frac{N}{2} + \mu)\phi} \ket{\mu},
\eeq 
where $\ket{\mu}$ are eigenstates of $\hat{J}_z$ and $s = (\sin\theta \cos \phi; \sin \theta \sin \phi; \cos\theta)$ is the mean spin direction.
We will focus on the dynamical evolution of the initial state $s_{x}=1$ ($s_y = s_z = 0$).
This corresponds to the coherent spin state given with all $N$ spins pointing along the $+x$ direction in the Bloch sphere. For the initial state, 
the collective spin mean values and variances are $\langle\hat{J}_x\rangle={N}/{2}$ and $\langle\hat{J}_\perp\rangle= 0$, $(\Delta\hat{J}_x)^2 = 0$ and $(\Delta\hat{J}_\perp)^2 ={N}/{4}$, respectively, 
where $\perp$ indicates an arbitrary direction orthogonal to the mean spin direction. 

We characterize the quantum dynamics by studying the QFI for unitary parametric transformations generated by an operator $\hat{J}$~\cite{CavesPRL1994, TothJPA2014, PezzeRMP2018}.
The QFI of a generic state $\hat{\rho}$ 
is formally defined as the variance $F_Q[\hat{\rho}, \hat{J}_n] = (\Delta \hat{L})^2$ of the symmetric logarithmic derivative $\hat{L}$, 
satisfying $2[\hat{J}, \hat{\rho}] = \{ \hat{L}, \hat{\rho}\}$~\cite{CavesPRL1994, TothJPA2014, PezzeRMP2018}. 
In terms of the spectral decomposition of $\hat{\rho}_t$ at time $t$
(namely $\hat{\rho}_t = \sum_\lambda p_\lambda \bra{\lambda} \ket{\lambda}$, with $p_\lambda \geq 0$ and $\ket{\lambda}$, eigenvalues and eigenvectors of $\hat{\rho}$, respectively), the QFI reads~\cite{CavesPRL1994, TothJPA2014, PezzeRMP2018}
\begin{equation} \label{QFIgeneral}
    F_Q(t) = 2 \sum_{\lambda,\lambda'} 
    \frac{(p_\lambda - p_{\lambda'})^2}{p_\lambda + p_{\lambda'}} \vert \langle \lambda \vert \hat{J} \vert \lambda' \rangle \vert^2.
\end{equation}
It has been shown~\cite{HaukeNATPHY2016} that Eq. (\ref{QFIgeneral}) can be related to the dynamical susceptibility, see also \cite{GabbrielliSCIREP2018,FrerotNATCOMM2019}.
For a pure quantum state $\ket{\psi_t}$ at time $t$,  Eq.~(\ref{QFIgeneral}) reduces to 
\begin{equation}
    F_Q(t) = 4(\Delta\hat{J})^2,
\end{equation}
where $(\Delta\hat{J})^2 = \bra{\psi_t} \hat{J}^2 \ket{\psi_t} - \bra{\psi_t} \hat{J} \ket{\psi_t}^2$.
On the experimental side, the relevant quantity to detect is Eq.~(\ref{QFIgeneral}). The variance of the operator $\hat{J}$ captures the behaviour of the QFI in the limit where the dynamics generating the quantum state is characterized by vanishing noise.

In the following we will restrict to the calculations of the QFI for pure states and for the operators $\hat{J} = \hat{J}_{x,y,z}$ introduced above.
In particular, the optimization of the QFI over $\hat{J}_{x,y,z}$ can be obtained as the 
maximum eigenvalue of the $3\times3$ covariance matrix
\begin{equation}
    [\mathbf{\Lambda}_Q]_{kl}=2[(\langle\hat{J}_k\hat{J}_l\rangle+\langle\hat{J}_l\hat{J}_k\rangle)- 2\langle\hat{J}_k\rangle\langle\hat{J}_l\rangle],
\end{equation}
with $k, l= x, y, z$, and the optimal direction is given by the corresponding eigenvector. 

The QFI is related to multipartite entanglement~\cite{note_multipartite, TothPR2009,  PezzePRL2009, HyllusPRA2012, TothPRA2012, TothJPA2014, PezzePNAS2016, PezzeRMP2018}:
a QFI satisfying $F_Q(t) > \mathcal{s}k^2+\mathcal{r}^2$, where $\mathcal{s} =\left\lfloor{N}/{k}\right\rfloor$ (being the maximum integer smaller than or equal
to ${N}/{k}$) and $\mathcal{r} = N- \mathcal{s}k$,  is a witness of $(k+1)$-partite entanglement among the $N$ bosonic particles in the state $\ket{\psi_t}$.
Multipartite entanglement generated the dynamical evolution of an  initially coherent spin state via Eq.~(\ref{TMTDH}) has been largely studied in the literature.
Theoretical works~\cite{PezzePRL2009, FerriniPRA2011} have focused on the one-axis twisting (OAT) Hamiltonian, corresponding to $a(t)=0$ in Eq.~(\ref{TMTDH})~\cite{Kitagawa1993, SorensenNATURE2001}.
The OAT dynamics has been realized experimentally by exploiting interatomic interaction in BECs \cite{Gross2010, Riedel2010, Schmied2016} 
or the atom-light interactions in cold thermal atoms \cite{Leroux2010,Takeuchi2005,Schleier-Smith2010, BohnetNATPHOT2014, HostenNATURE2016}, 
see Ref.~\cite{PezzeRMP2018} for a comprehensive review.
Another model of interest is the twist-and-turn (TAT) Hamiltonian, corresponding to 
$a(t) = A > 0$ in Eq.~(\ref{TMTDH})~\cite{Milburn1997, Cirac1998, Steel1998, MicheliPRA2003}.
The TAT dynamics is characterized by the competition between 
the twisting (due to the quadratic term in the Hamiltonian) and the linear coupling between the two modes that turns the state along an orthogonal direction. 
It has been shown, both experimentally \cite{StrobelSCIENCE2014, Muessel2015} and theoretically \cite{SorelliPRA2019} that the TAT model creates entanglement faster than the OAT model.
More recently, multipartite entanglement witnessed by the QFI has been studied for chaotic systems, where $a(t)$ has a $\delta$-function time dependence~\cite{LerosePRA2020b}.

Furthermore, as shown in Refs.~\cite{Lewis-Swan2019, GarrtnerPRL2018} the QFI can be also directly connected to 
the out-of-time-order correlations. 
Let us introduce the quantity
$\mathcal{F}_t(\phi) = {\rm tr}[\hat{\rho}_0 \hat{\rho}_t(\phi)] = 
{\rm tr}[\hat{W}_t^\dag(\phi) \hat{\rho}_0^\dag \hat{W}_t(\phi) \hat{\rho}_0]$,
where $\hat{\rho}_0$ is the initial state, 
$\hat{\rho}_t(\phi) = e^{i \hat{H} t} e^{-i \hat{J} \phi} e^{-i \hat{H} t} \hat{\rho}_0 e^{i \hat{H} t} e^{i \hat{J} \phi} e^{-i \hat{H} t}$, and $\hat{W}_t(\phi) = e^{i \hat{H} t} e^{-i \hat{J} \phi} e^{-i \hat{H} t}$ and $\hat{W}^\dag_t(\phi) = e^{i \hat{H} t} e^{i \hat{J} \phi} e^{-i \hat{H} t}$.
Notice that, in the definition above we have considered a time-independent Hamiltonian $\hat{H}$, as relevant in our case (see discussion in Sec. III).
Expanding in Taylor series for $\phi \ll 1$, we have 
$\mathcal{F}_t(\phi) = {\rm tr}[\hat{\rho}_t^2] - \phi^2 {\rm tr}[\hat{\rho}_t^2 \hat{J}^2 - (\hat{\rho}_t \hat{J})^2] + O(\phi^3)$, 
where $\hat{\rho}_t = e^{i \hat{H} t} \hat{\rho}_0 e^{-i \hat{H} t}$.
The second derivative with respect to $\phi$ is~\cite{GirolamiENTROPY2017, GarrtnerPRL2018, Lewis-Swan2019}
\be
- 2\frac{d^2 \mathcal{F}_t(\phi)}{d\phi^2} = 4 {\rm tr}[\hat{\rho}_t^2 \hat{J}^2 - (\hat{\rho}_t \hat{J})^2] \leq F_Q(t),
\ee
with equality for pure states $\hat{\rho}_0 = \ket{\psi_0} \bra{\psi_0}$.
It should be noticed that, for pure states, $\mathcal{F}_t(\phi)$ writes as an out-of-time-order correlator, 
$\mathcal{F}_t(\phi) = \mean{\hat{W}_t^\dag(\phi) \hat{V}_0^\dag \hat{W}_t(\phi) \hat{V}_0}$ for $\hat{V}_0 = \ket{\psi_0} \bra{\psi_0}$ and the mean value calculated on the initial state $\ket{\psi_0}$. 
%

While exact numerical calculations of the QFI for the model (\ref{TMTDH}) can be obtained for long times and large systems, it is interesting to study various approximations to the quantum dynamics of the QFI. 
For instance, Ref.~\cite{SorelliPRA2019} has considered an exact quantum phase model,  
Refs.~\cite{Julia-DiazPRA2012a, Julia-DiazPRA2012b} have used the a Fock state expansion~\cite{JavanainenPRA1999, ShchesnovichPRA2008}
for the study of the spin-squeezing parameter, which is a lower bound to the QFI~\cite{PezzeRMP2018}, 
while Ref.~\cite{LerosePRA2020b} has proposed a semiclassical approximation.
In this paper, we study the time evolution of the QFI within a BMF approach~\cite{AnglinPRA2001b}. 

Here, rather than using a Schr\"odinger picture, we consider the dynamics of the spin operators. 
According to the Heisenberg equation for the operator $\hat{J}_l$, $i({d\hat{J}_l}/{dt})=[\hat{J}_l,\hat{H}(t)]$, the expectation value of $\hat{J}_l$
depends not only on themselves, but also on the second-order moments $\langle\hat{J}_l\hat{J}_k\rangle$. Similarly, the time evolution of the second-order
moments depends on third-order moments, and so on.
Consequently, we can obtain the hierarchy of equations
of motion for the expectation values $\langle\hat{J}_l\rangle$, $\langle\hat{J}_l\hat{J}_k\rangle$, $\dots$. The second-order truncation of the hierarchy equations and the approximation $\langle\hat{J}_l\hat{J}_k\hat{J}_j\rangle\simeq\langle\hat{J}_l\hat{J}_k\rangle\langle\hat{J}_j\rangle+\langle\hat{J}_l\rangle\langle\hat{J}_k\hat{J}_j\rangle+\langle\hat{J}_l\hat{J}_j\rangle\langle\hat{J}_k\rangle-2\langle\hat{J}_l\rangle\langle\hat{J}_k\rangle\langle\hat{J}_j\rangle$ will give the equations of motion of the expectation values for the first- and second-order moments, which are \cite{AnglinPRA2001b}
\begin{widetext}
\begin{equation}\label{beyondmean-fieldE}
\frac{d}{dt}\left(\begin{array}{c}
s_x\\s_y\\s_z\\\Delta_{xz}\\\Delta_{yz}\\\Delta_{xy}\\\Delta_{xx}\\\Delta_{yy}\\\Delta_{zz}\\
\end{array}\right)
=\left(\begin{array}{ccccccccc}
 0   & -cs_z  & 0     & 0        & -\frac{c}{2} & 0 & 0 & 0 & 0\\
cs_z & 0      & -a(t) & \frac{c}{2} & 0 & 0 & 0 & 0 & 0 \\
0 & a(t) & 0 & 0 & 0 & 0 & 0 & 0 & 0\\
0 & 0 & 0 & 0 & -cs_z & a(t) & 0 & 0 & -cs_y\\
0 & 0 & 0 & cs_z & 0 & 0 & 0 & a(t) & cs_x-a(t)\\
0 & 0 & 0 & cs_x-a(t) & -cs_y & 0 & cs_z & -cs_z & 0\\
0 & 0 & 0 & -2cs_y & 0 & -2cs_z & 0 & 0 & 0\\
0 & 0 & 0 & 0 & 2(cs_x-a(t)) & 2cs_z & 0 & 0 & 0\\
0 & 0 & 0 & 0 & 2a(t) & 0 & 0 & 0 & 0\\
\end{array}\right)\left(\begin{array}{c}
s_x\\s_y\\s_z\\\Delta_{xz}\\\Delta_{yz}\\\Delta_{xy}\\\Delta_{xx}\\\Delta_{yy}\\\Delta_{zz}\\
\end{array}\right),   
\end{equation}
\end{widetext} 
where $(s_x,s_y,s_z)={2}(\langle\hat{J}_x\rangle,\langle\hat{J}_y\rangle,\langle\hat{J}_z\rangle)/{N}$ and $\Delta_{lk}={4}(\langle\hat{J}_l\hat{J}_k+\hat{J}_k\hat{J}_l\rangle-2\langle\hat{J}_l\rangle\langle\hat{J}_k\rangle)/{N^2}$. This BMF model is also referred to as the Bogoliubov
backreaction model \cite{AnglinPRA2001b, LerosePRL2018}. Taking ${N^2}\Delta_{lk}/{8}$ as the matrix elements can also form a similar covariance matrix $\mathbf{\Lambda}_B$ with $\Delta_{lk}=\Delta_{kl}$ and four times the maximum eigenvalue gives the QFI within the BMF approximation, namely, $F_B$.

\section{Benchmark models}

While the main focus of this paper is on the QKR model, 
which is discussed in the next section, we first analyze here the QFI calculated within the BMF approach for the OAT and TAT models.

\subsection{One-axis-twisting dynamics}

\begin{figure}[t]
\includegraphics[width=\linewidth]{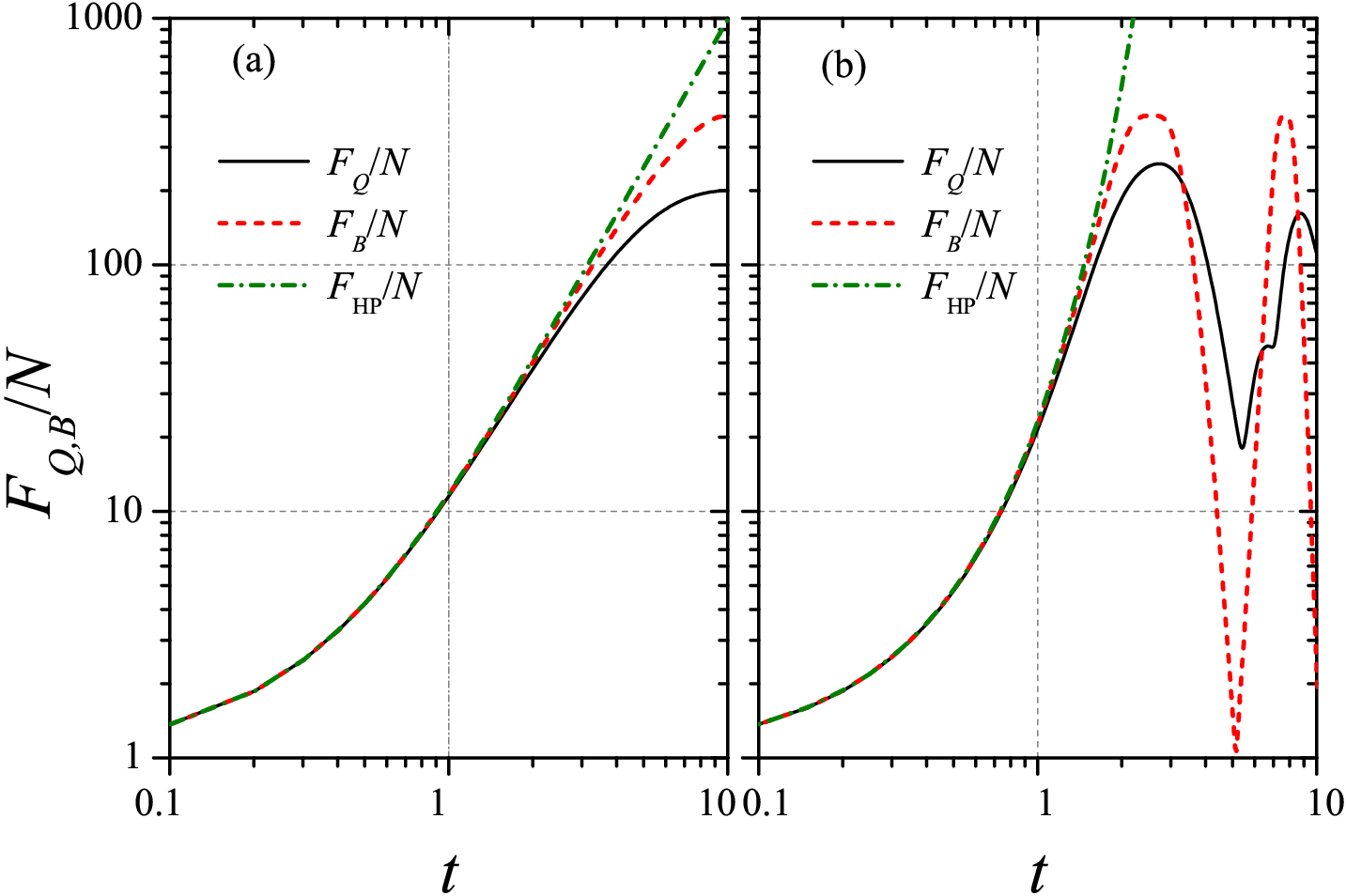}
\caption{QFI dynamics for (a) the OAT model [$a(t)=0$ in Eq.~(\ref{TMTDH})] and (b) the TAT model [for $A={\pi}/{2}$]. The solid black line is the exact solution (see text), the dashed line is the BMF approach $F_B(t)$, while the $F_{\rm HP}$ is the semiclassical HP Eq.  (\ref{HPOATjiexi}). We have set $c=\pi$, $N=400$ and an initial coherent spin state pointing along the $x$ axis ($s_x=1$).}
\label{newfig8}
 \end{figure}

For the OAT model, it is possible to obtain an analytical expression for the QFI~\cite{Kitagawa1993, PezzePRL2009, PezzeRMP2018, note1} at all times and for arbitrary number of particles.  
For times $ct \leq N\pi/2-2\sqrt{N}$, 
the QFI is given by 
\begin{equation}\label{FQjiexi}
\frac{F_Q(t)}{N}=1+\frac{(N-1)}{4}\big[\alpha_Q(t)+\sqrt{\alpha_Q^2(t)+\beta_Q^2(t)}\big],
\end{equation}
where $\alpha_Q(t)=1-\cos^{N-2}({2ct}/{N})$ and $\beta_Q(t)=4\sin({ct}/{N})\cos^{N-2}({ct}/{N})$. 
The short-time evolution of the QFI is characterized by a sudden increase starting from $F_Q(0)=N$, see Fig.~\ref{newfig8}. 
We consider $N \gg 1$, such that $N-1\simeq N-2 \simeq N$.
A Taylor expansion gives
\begin{equation}\label{FQzhankai}
\frac{F_Q(t)}{N} = 1 + c t + \frac{c^2t^2}{2} +\frac{c^3 t^3}{8} - \frac{3c^3 t^3}{4N}+O\big(c^4t^4\big).    
\end{equation}
Notice that, up to the second order in $ct$, the Taylor expansion is analogous to that of $e^{ct}$.
For times $ct \geq 1$ we have $\alpha_Q(t) \gg \beta_Q(t)$ and Eq.~(\ref{FQjiexi}) becomes
\be \label{FQapp}
\frac{F_Q(t)}{N} \simeq 1+\frac{N\alpha_Q(t)}{2}.
\ee
For times $1 \leq ct \leq 2\sqrt{N}$ we have $\alpha_Q(t) \simeq 2 c^2 t^2/N$ such that 
\be \label{FQappt2}
\frac{F_Q(t)}{N} \simeq c^2 t^2,
\ee
having a quadratic scaling with time. 
For  $ct \geq 2\sqrt{N}$, we have $\cos^{N-2}({2ct}/{N}) \ll 1$ and thus obtain $\alpha_Q(t) \simeq 1$.
Equation (\ref{FQapp}) thus predicts a constant plateau $F_Q(t)=N^2/2$ as a function of time, up to $ct \leq N\pi/2 - 2\sqrt{N}$.
Finally, for $ct \geq N\pi/2 - 2\sqrt{N}$ the QFI further increases and reaches the maximum value $F_Q(t) = N^2$ at time $ct = N\pi/2$. It should be mentioned that the behavior of QFI given by Eq. (\ref{FQjiexi}) is obviously $\pi N$-periodic. 

We compare the exact QFI to that calculated within the BMF approach, which can also be solved analytically.
For the initial state $s_x(0)=1$, the covariance matrix $\mathbf{\Lambda}_B$ reads
\begin{equation}\label{BMFCVM}
\mathbf{\Lambda}_B(t)=\frac{N^2}{8}\left(\begin{array}{ccc}
0 & 0 & 0 \\
0 & \frac{2+N(1-\cos\frac{2ct}{\sqrt{N}})}{N} & \frac{2\sin\frac{ct}{\sqrt{N}}}{\sqrt{N}}\\
0 & 0 & \frac{2}{N}\\
\end{array}\right).
\end{equation}
The QFI is given by four times the maximum eigenvalue of $\mathbf{\Lambda}_B(t)$, namely
\begin{equation}\label{FBjiexi}
\frac{F_B(t)}{N}=1+\frac{N\alpha_B(t)+\sqrt{8N\alpha_B(t)+N^2\alpha_B^2(t)}}{4}, 
\end{equation}
where $\alpha_B(t)=2\sin^2({c t}/{\sqrt{N}})$. 
Expanding Eq. (\ref{FBjiexi}) in a Taylor series for ${ct}/{N} \ll 1$, we obtain
\begin{equation}\label{FBzhankai}
\frac{F_B(t)}{N} = 1 + c t + \frac{c^2t^2}{2} +\frac{c^3 t^3}{8} - \frac{c^3 t^3}{6N}+O\big(c^4t^4\big).
\end{equation}
Equation~(\ref{FBzhankai}) coincides with Eq.~(\ref{FQzhankai}) up to the second order in $ct$. 
The difference in the third order is the term $7 c^3 t^3 /12N$, which is negligible asymptotically for large $N$.
Similar considerations can be obtained for higher order terms in the Taylor expansion.
For time $ct \geq 1$ we have $\alpha_B^2(t) \gg \alpha_B(t)$ and Eq.~(\ref{FBjiexi}) becomes
\be
\frac{F_B(t)}{N} \simeq 1+\frac{N\alpha_B(t)}{2},
\ee
in analogy with Eq.~(\ref{FQapp}).
A Taylor expansion for $ct/\sqrt{N} \gg 1$, predicts the quadratic scaling
\be \label{FBappt2}
\frac{F_B(t)}{N} \simeq c^2 t^2,
\ee
for $1 \leq ct \leq \sqrt{N}$, in agreement with Eq.~(\ref{FQappt2}).
For longer times, Eq.~(\ref{FBjiexi}) is characterized by harmonic oscillations of period $\sqrt{N}\pi/c$.
At half period, namely, $t={m\sqrt{N}\pi}/{2c}$ with $m=1,3,\dots$, 
$\alpha_B =  2N$, Eq.~(\ref{FBjiexi}) (wrongly) predicts the Heisenberg limit 
$F_B \simeq N^2$, see Fig.~\ref{newfig8}.
We thus conclude that the BMF approach fails for times $ct \geq \sqrt{N}$.

A different approximation to the exact QFI dynamics has been recently studied in Refs.~\cite{LerosePRR2020a, LerosePRA2020b} 
using the Holstein-Primakoff (HP) transformation
around the instantaneous mean spin polarization (lying on the $x$ axis, in the present case).
Hence, the HP transformation is performed by mapping
the transverse ($z$ and $y$) spin components to canonical bosonic variables
\begin{equation}\label{HPexpand}
\hat{J}_x =\frac{N}{2}-\hat{n},~~~\hat{J}_y\simeq\sqrt{\frac{N}{2}}\hat{p},~~~\hat{J}_z\simeq\sqrt{\frac{N}{2}}\hat{q},
\end{equation}
where the bosonic variables $\hat{q}$ and $\hat{p}$ satisfy $[\hat{q},\hat{p}]=i$, $\hat{n}=({\hat{q}^2+\hat{p}^2-1})/{2}$. 
The quantum number $n$ ($= 0, 1, 2,\dots$) labels the quantized spin projection along the $x$ direction  of the classical minimum. 
Using Eqs. (\ref{HPexpand}), the Hamiltonian (\ref{TMTDH}) can be approximated by neglecting terms of order ${1}/{N}$,
\begin{equation}\label{simcH}
\hat{H}(t)\simeq a(t)\left(\frac{N}{2}-\hat{n}\right)+c\frac{\hat{q}^2}{2},
\end{equation}
and thus one can easily obtain the equations of motion for
the quantum fluctuations 
\begin{equation}\label{QFDME}
\frac{d\hat{q}}{dt}=-a(t)\hat{p},~~~~\frac{d\hat{p}}{dt}=[a(t)-c]\hat{q}.    
\end{equation}
This evolution of the quantum fluctuations is exact at the quadratic level and corresponds to the classical one. We start from initial coherent spin state $s_x(0)=1$, for which $\langle\hat{q}^2(0)\rangle = \langle\hat{p}^2(0)\rangle = {1}/{2}$ and
$\langle\hat{q}(0)\hat{p}(0)\rangle = \langle\hat{p}(0)\hat{q}(0)\rangle = 0$. 
As a result, the QFI can be given by~\cite{LerosePRR2020a, LerosePRA2020b} 
\begin{equation}\label{HPOATjiexi}
\frac{F_{\rm HP}(t)}{N}=1+2\langle\hat{n}_\mathrm{exc}(t)\rangle+2\sqrt{\langle\hat{n}_\mathrm{exc}(t)\rangle(\langle\hat{n}_\mathrm{exc}(t)\rangle+1)},    
\end{equation}
where $\langle\hat{n}_\mathrm{exc}(t)\rangle=({\langle\hat{q}^2\rangle+\langle\hat{p}^2\rangle-1})/{2}$ is the number of collective excitations. 
For the OAT model, $\langle\hat{n}_\mathrm{exc}(t)\rangle=c^2t^2/4$. The Taylor expansion at short times
leads to~\cite{LerosePRR2020a,LerosePRA2020b,LerosePRIVATE}
\begin{equation}\label{HPOATzhankai}
\frac{F_{\rm HP}(t)}{N} =  1+ct+\frac{c^2t^2}{2}+\frac{c^3t^3}{8}+O(c^5t^5).  
\end{equation}
The HP approximation is expected to hold for time sufficiently long as far as the mean spin length is $\mean{\hat{J}_x} \simeq N/2$, 
namely, as far as we can locally approximate the generalize ($N$-spin) Bloch sphere with the tangent plane~\cite{PezzeRMP2018}.
Using the exact result $\mean{\hat{J}_x} = (N/2) \cos^{N-1} (ct)$~\cite{Kitagawa1993}, we obtain the HP approximation
is expected to hold up to times $ct = O(\sqrt{N})$.
At intermediate times, for $1 \ll ct \ll \sqrt{N}$, we can approximate 
\be
\frac{F_{\rm HP}(t)}{N} \simeq c^2 t^2,
\ee
predicting a quadratic scaling of the QFI with time, in agreement with Eqs.~(\ref{FQappt2}) and (\ref{FBappt2}).

In Fig. \ref{newfig8}(a), we compare the exact $F_Q(t)$,  Eq. (\ref{FQzhankai}), the BMF approximation $F_{B}(t)$, Eq. (\ref{FBzhankai}), and the 
HP approximation, Eq.~(\ref{HPOATzhankai}), as a function of time, up to $ct \leq \sqrt{N}$. 
As discussed above, we see that all models agree very well for short times, see also~Refs. \cite{SorelliPRA2019, Julia-DiazPRA2012b}.
We also see that, as expected, the BMF approach is closer than the HP approximation to the exact solution. 
This is mainly because the BMF approximation (\ref{beyondmean-fieldE}) considers both the coupling 
between quantum fluctuations and the mean spin value and the coupling between fluctuations, 
while the semiclassical method based on HP transformation only considers the coupling between quantum fluctuations. 

\subsection{Twist-and-turn dynamics}

The exact dynamics for $F_Q(t)$ is not analytical, except for weak interaction~\cite{LawPRA2001}.
The most interesting situation is that corresponding to the unstable regime~\cite{SorelliPRA2019} (i.e., $c>A$).
The quantity $F_{B}(t)$ is obtained by diagonalizing the covariance matrix $\mathbf{\Lambda}_{B}$ on the basis of solving Eq.~(\ref{beyondmean-fieldE}) numerically. 
For short time, we can obtain an analytical solution when taking $s_x(t)\simeq 1-(A^2+\lambda^2)^2t^2/2NA^2$ with $\lambda^2=A(c-A)$. The Taylor expansion reads
\beq\label{TATFBzhankai}
\frac{F_B(t)}{N}=&&1+ct+\frac{c^2t^2}{2}+\frac{c\lambda^2t^3}{6}+\frac{c^3t^3}{8}-\frac{c^3t^3}{2N}\nonumber\\&&+\frac{c^2\lambda^2t^4}{6}-\frac{c^4t^4}{2N}+O(c^5t^5),
\eeq
that agrees with Ref.~\cite{SorelliPRA2019}.
Similarly to the OAT~\cite{Julia-DiazPRA2012b}, the Taylor expansion up to the second order is analogous to that of $e^{ct}$.
Nevertheless, the third order of Eq.~(\ref{TATFBzhankai}) is larger than that of Eq.~(\ref{FBzhankai})
by a factor ${\lambda^2}/{c^2} =  {A}( 1 - {A}/{c})/{c}$, 
that has a maximum at $c/A = 2$~\cite{SorelliPRA2019}.
We can thus conclude that the BMF approximation breaks down for times of the order of
\be \label{TATtc}
t_c \simeq \frac{1}{\lambda} \log \bigg( \frac{N \lambda^2}{cA} \bigg),
\ee
where $t_c$ has been identified in Ref.~\cite{MicheliPRA2003} as the time needed to generate a macroscopic superposition state.

We can compare this prediction with the semiclassical HP approach of Ref.~\cite{LerosePRA2020b}.
The analytical solution of $F_{\rm HP}(t)$ is still given by Eq. (\ref{HPOATjiexi}) where, in the unstable regime,
$\langle\hat{n}_\mathrm{exc}(t)\rangle=({A}/{\lambda}+{\lambda}/{A})^2\sinh^2(\lambda t)/{4}$ \cite{LerosePRIVATE}.
The Taylor expansion of $F_{\rm HP}(t)$ at short times reads \cite{LerosePRR2020a,LerosePRA2020b,LerosePRIVATE}
\begin{equation}\label{HPTATzhankai}
\frac{F_{\rm HP}(t)}{N}= 1+ct+\frac{c^2t^2}{2}+\frac{c\lambda^2t^3}{6}+\frac{c^3t^3}{8}+\frac{c^2\lambda^2t^4}{6}+O(c^5t^5).
\end{equation}
Similarly to the OAT case, the agreement between Eqs.~(\ref{TATFBzhankai}) and (\ref{HPTATzhankai}) is up to the second order of $ct$.  
The difference in the third order term is $c^3 t^3/2N$, which is negligible asymptotically for large $N$.
Similar considerations can be obtained for higher order terms in the Taylor expansion.
Furthermore, neglecting subleading terms for $N \gg 1$, Eqs.~(\ref{TATFBzhankai}) and (\ref{HPTATzhankai}) agree with the results of Ref.~\cite{SorelliPRA2019}.
Finally, in the limit $\langle\hat{n}_\mathrm{exc}(t)\rangle \gg 1$, that is obtained after a transient time, $c / \lambda \leq ct \leq \log (N \lambda^2/cA)$, 
the HP approach predicts~\cite{LerosePRA2020b,LerosePRIVATE} 
\be
\frac{F_{\rm HP}(t)}{N} \simeq \bigg( \frac{A}{\lambda}+ \frac{\lambda}{A} \bigg) e^{2 \lambda t},
\ee
where the largest instability parameter $\lambda/c$ is obtained for $c/A=2$. 
A comparison between the exact $F_Q(t)$ and the approximations $F_B(t)$ and $F_{\rm HP}(t)$ is shown in Fig.~\ref{newfig8}(b).

\section{Quantum kicked rotor model}

In the following we consider the case
\be \label{at}
a(t)=\frac{AT}{\tau_1}\sum_{n=0}^{\infty}\delta(t-\tau_n), 
\ee
such that the two-mode time-dependent Hamiltonian (\ref{TMTDH}) becomes
\begin{equation}\label{QKR}
\hat{H}(t)=\frac{AT}{\tau_1}\sum_{n=0}^{\infty}\delta(t-\tau_n)\hat{J}_x+\frac{c}{N}\hat{J}_z^2,
\end{equation}
where $A$ is the strength of the kicked field. The Dirac $\delta$-function presents a rectangular-pulse-type kicked field with $\tau_n\in[nT,nT+\tau_1]$. Here, $T=\tau_0+\tau_1$ is the period of the kicked field with $\tau_0$ and $\tau_1$ being the interval and the width of the kicks, respectively. Equation (\ref{QKR}) corresponds to the QKR model~\cite{Haake1987, HaakeBOOK}.
The periodicity of the two-mode QKR model (\ref{QKR}) allows us to describe stroboscopically the evolution of quantum state with
discrete time in units of $T$, i.e.,
\begin{equation}
|\psi(n)\rangle=U_R(T)|\psi(n-1)\rangle=U_R^n(T)|\psi(0)\rangle,    
\end{equation}
with the unitary Floquet operator
\begin{equation}
U_R(T)={e}^{-i[\frac{c}{N}\hat{J}_z^2]\tau_0}{e}^{-i[\frac{AT}{\tau_1}\hat{J}_x+\frac{c}{N}\hat{J}_z^2]\tau_1}.   
\end{equation}
When $\tau_1\rightarrow 0$, $\tau_0\rightarrow T$, the kicked field  becomes a $\delta$-function-type form and the Floquet operator reduces to
\begin{equation}
U_\delta(T)={e}^{-i[\frac{c}{N}\hat{J}_z^2]T}{e}^{-i[{A}\hat{J}_x]T}.   
\end{equation}
In our calculations, we adopt a rectangle-pulse-type kicked field by setting $\tau_0=1$ and $\tau_1=0.01$, which is very close to the $\delta$-function-type kicked field.

\begin{figure}[t!]
\includegraphics[width=\linewidth]{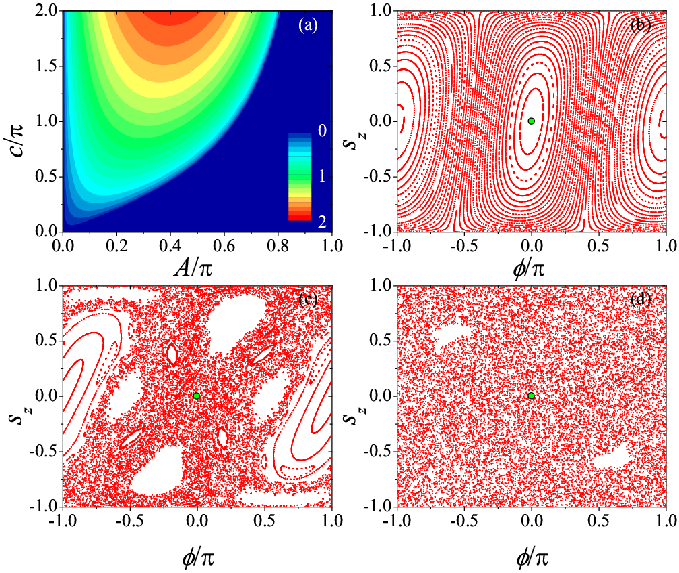}
\caption{(a) Maximum Lyapunov exponents for the initial state $s_x(0)=1$ [i.e., $(s_z,\phi)=(0,0)$]. (b), (c), and (d) show the Poincar\'{e} sections with $A/\pi=0.4$ for $c/\pi=0.2$, $0.8$, and $1.4$, respectively. Green solid  points indicate the initial states adopted in (a).}\label{fig1}
\end{figure}

\subsection{Mean-field approximation and chaotic motion}

Here, the total number $N$ is conserved and $1/N$ can be identified with an effective Planck constant.
The mean-field (MF) limit $N\rightarrow\infty$ (and thus the interaction $c/N\rightarrow 0$) corresponds to the classical limit, where
$(s_x,s_y,s_z)$ become classical variables that form a single-particle Bloch vector confined to the unit sphere and thus the norm is conserved \cite{AnglinPRA2001b}.
$(s_z,\phi)$ can be identified with classical phase-space variables, where $\phi$ is the azimuthal angle. The first-order truncation of the hierarchy equations together with the approximations  $\langle\hat{J}_l\hat{J}_k\rangle\simeq\langle\hat{J}_l\rangle\langle\hat{J}_k\rangle$ give the equations of motion of the expectation values for the first-order moments, namely, the MF equation \cite{AnglinPRA2001b}
\begin{equation}\label{mean-fieldE}
\frac{d}{dt}\left(\begin{array}{c}
s_x \\
s_y \\
s_z \\
\end{array}\right)=\left(\begin{array}{ccc}
 0 & -cs_z  & 0 \\
cs_z & 0 & -a(t) \\
0 & a(t) & 0 \\
\end{array}\right)\left(\begin{array}{c}
s_x \\
s_y \\
s_z \\
\end{array}\right),   
\end{equation}
where $a(t)$ is given by Eq.~(\ref{at}). 
For such a periodic-driven system, the classical motion can usually be divided into two classes: i) stable and periodic motion and ii) unstable and chaotic motion. In the unstable case, the magnification of the initially small deviation is exponential in time, which can be measured by the maximum Lyapunov
exponent, i.e., \cite{Liu2005} 
 \be
\lambda_L(A, c)=\lim_{m\rightarrow\infty}\frac{1}{mT}\sum_{n=1}^{m}\ln\frac{||(\delta s_x(nT),\delta s_y(nT),\delta s_z(nT))||}{||(\delta s_x(0),\delta s_y(0),\delta s_z(0))||}, 
\ee
where $\delta s_{x,y,z}(0)=|s'_{x,y,z}(0)-s_{x,y,z}(0)|$ denote the initial deviations and $\delta s_{x,y,z}(nT)=|s'_{x,y,z}(nT)-s_{x,y,z}(nT)|$ denote the deviations at $t=nT$, where $s_{x,y,z}(t)$ and $s'_{x,y,z}(t)$ respectively represent the two classical trajectories without perturbation and with perturbation initially. 
In Fig. \ref{fig1}(a) we plot the maximum Lyapunov
exponents in the $(A,c)$-plane for the initial state $s_x(0)=1$. In our numerical calculations we set $\delta s_{x}(0)=10^{-5}$ and evolve the system for 500 periods. 
For the stable motion, $\lambda_L$ is zero (dark blue region), while for the chaotic regime $\lambda_L$ is positive. For example, in Fig. \ref{fig1}(a) for $A/\pi=0.4$ and $c/\pi=0.2$, the initial state is in stable region, while $A/\pi=0.4$ both for $c/\pi=0.8$ and $c/\pi=1.4$ the initial states are in chaotic region. However, the point $(A,c)/\pi=(0.4,1.4)$ is ``more chaotic'' than $(A,c)/\pi=(0.4,0.8)$ is, because $(A,c)/\pi=(0.4,1.4)$ ($\lambda_L\simeq 1.524$) has a larger Lyapunov exponent than $(A,c)/\pi=(0.4,0.8)$ ($\lambda_L\simeq 0.966$), 
meaning that $\delta s_{x,y,z}(nT)$ deviate faster from the initial conditions.

To show the property of classical motion of the kicked system, we plot the Poincar\'{e} sections by using the stroboscopic points at $t=nT$ in Figs. \ref{fig1}(b)-\ref{fig1}(d) for different interaction strengths $c/\pi=0.2$, $0.8$, and $1.4$ with same kicked strength $A/\pi=0.4$. We see that for the integrable situation [e.g., $(A,c)/\pi=(0.4,0.2)$, see Fig. \ref{fig1}(b)], all trajectories (periodic orbit and elliptic fixed point) are stable with the maximum Lyapunov
exponents being zero. In the completely nonintegrable case [e.g., $(A,c)/\pi=(0.4,1.4)$, see Fig. \ref{fig1}(d)], we see that the phase space is full of chaotic trajectories with largely positive $\lambda_L$. In the mixed case [e.g., $(A,c)/\pi=(0.4,0.8)$, see Fig. \ref{fig1}(c)], the phase space illustrates one big-size island, six middle-size islands, and four small-size islands (see the blank regions). It is clear that the motions inside the islands are stable and the motions outside
the islands are unstable and mainly chaotic with medium positive $\lambda_L$.
In Figs. \ref{fig1}(b)-\ref{fig1}(d), the green points mark the initial state $s_{x}(0)=1$. 

\subsection{Quantum Fisher Information and maximum Lyapunov exponent}

\begin{figure}[t]
\includegraphics[width=\linewidth]{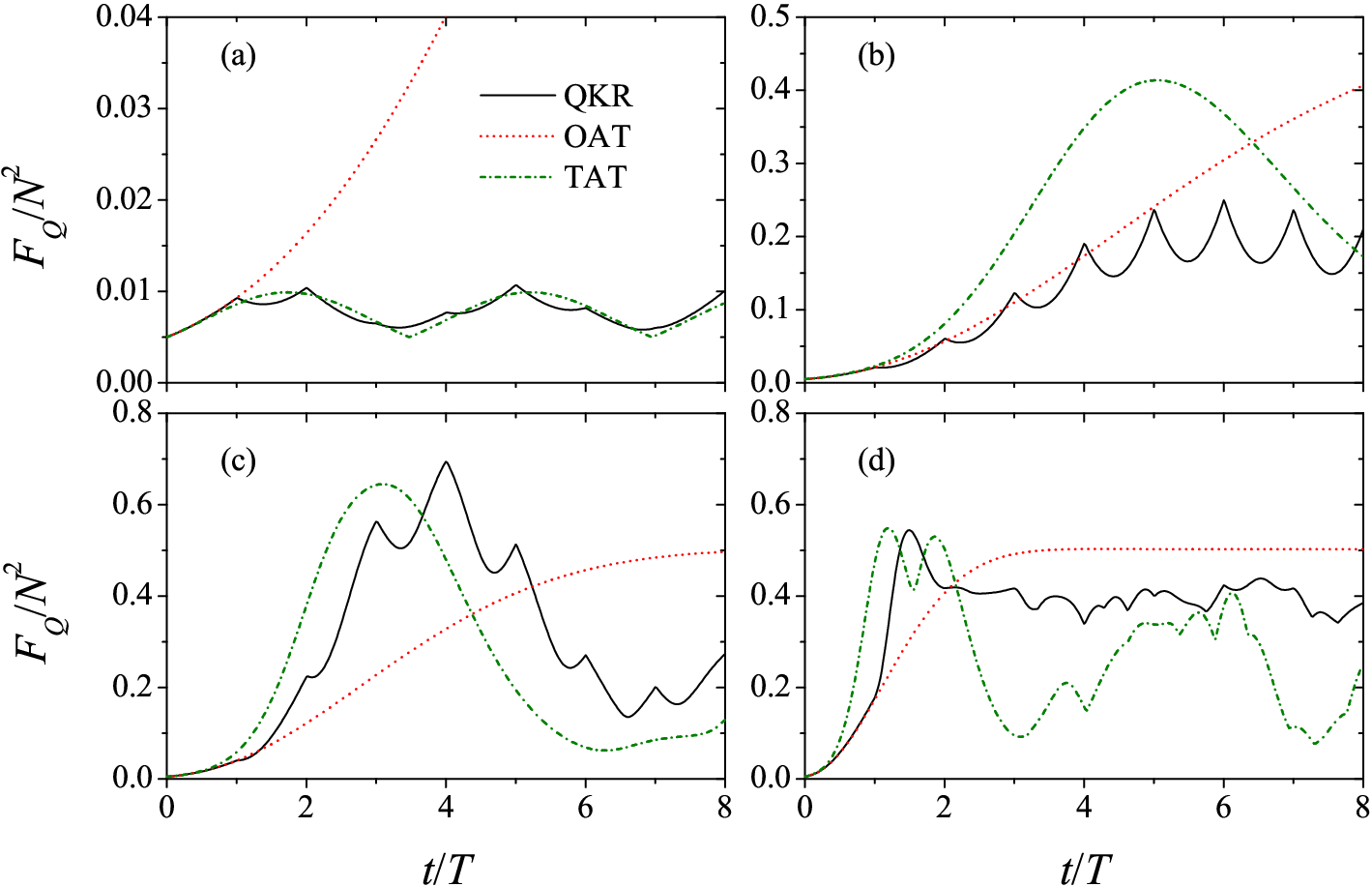}
\caption{Time evolution of QFI for the QKR ($A/\pi=0.4$, black solid lines), the OAT ($A=0$, red dashed lines), and the TAT ($A/\pi=0.4$, green dash-dotted lines) models with different interaction strengths: (a) $c/\pi=0.2$ (the initial state being a stable point), (b) $c/\pi=0.5$ (the initial state being a saddle point), (c) $c/\pi=0.8$ [The initial state is in chaotic regime, see Fig. \ref{fig1}(c)], and (d) $c/\pi=2.0$ (The initial state is in deep chaotic regime). We have set $N=200$ and $s_x(0)=1$.} 
\label{newfig7}
 \end{figure}

To show the influence of the chaotic motions on the dynamics of entanglement, 
we calculate the QFI for the same initial state [$s_x(0)=1$] and different interacting parameters. 
In Fig. \ref{newfig7}, for a fixed value of the parameter $c$, we compare the time-evolution of the QFI between the QKR model
(solid black line; obtained for parameter $A$ with $\delta$-kicked time dependence), 
the OAT model (dashed red line; $A=0$), 
and the TAT model (dot-dashed green line; for constant $A$). 
When the initial quantum state corresponds to a stable point (i.e., $c/\pi=0.2$) or a saddle point (i.e., $c/\pi=0.5$) in the Poincar\'{e} section of the classical model, the QKR model does not show any advantage in entanglement generation  because of the small oscillation [see Fig. \ref{newfig7}(a)] or the slow growth [see Fig. \ref{newfig7}(b)] of the QFI with time. However, when the initial quantum state is located in the classical chaotic region (i.e., $c/\pi=0.8$, and 2.0), compared with the QFI for the OAT model, the  QFI for the QKR model can reach and exceed ${N^2}/{2}$ in a very short period of time (e.g., 2 or 3 kicks), which indicates that the quantum chaos can significantly accelerate the generation of entanglement [see Fig. \ref{newfig7}(c) and \ref{newfig7}(d)]. 
When compared with the TAT model, the QFI for the QKR model is slower in all cases.
In particular, it was shown in Ref.~\cite{SorelliPRA2019} that the maximally unstable TAT dynamics (for $c/A=2$) is 
characterized by the fastest generation of entanglement within quadratic Hamiltonians as in Eq.~(\ref{TMTDH}).
 
\begin{figure}[b!]
\includegraphics[width=\linewidth]{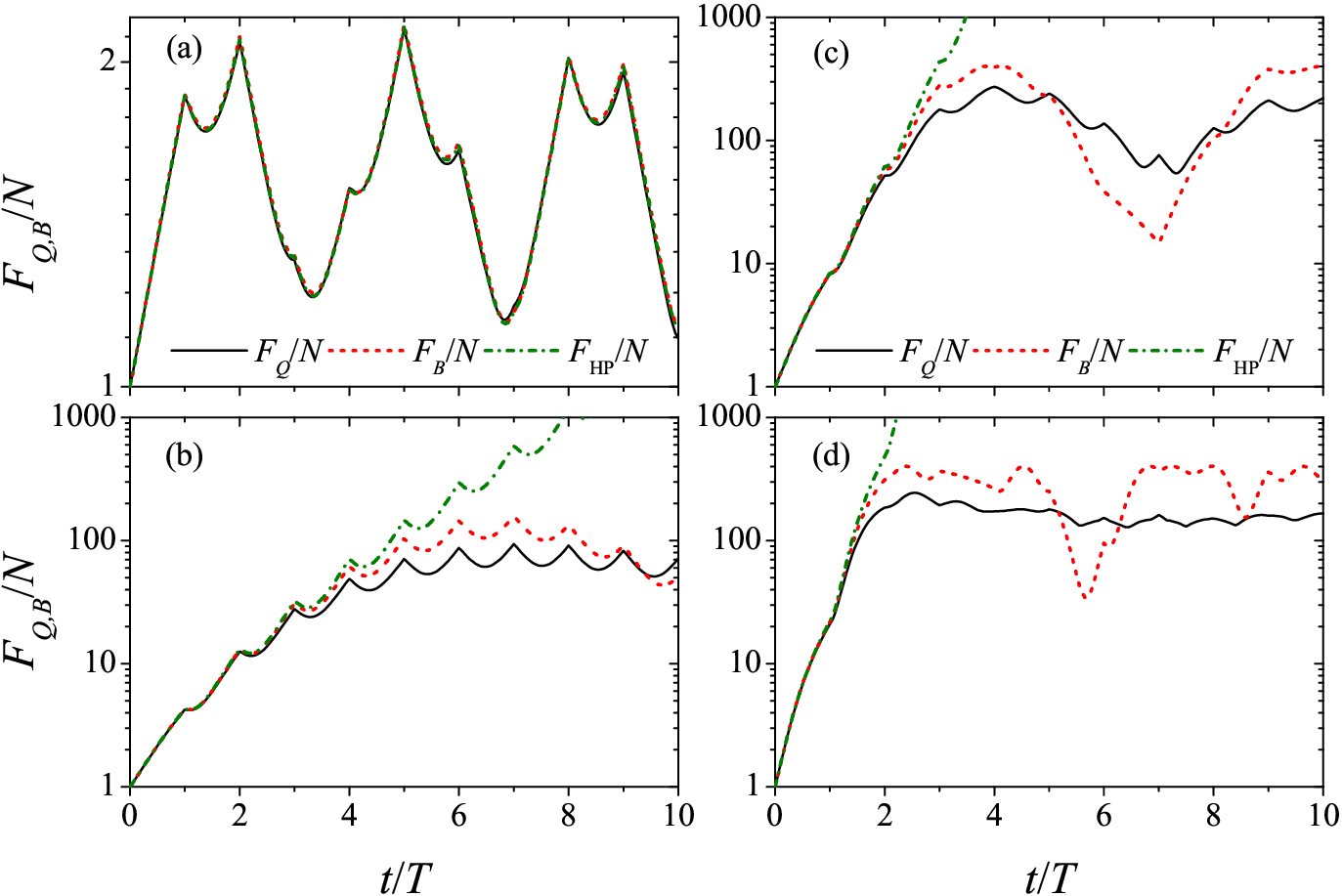}
\caption{Comparison of QFI dynamics for the QKR model with the initial state $s_x(0)=1$ being (a) a stable fixed point (i.e., $c/\pi=0.2$), (b) a saddle point (i.e., $c/\pi=0.5$), (c) in chaotic regime (i.e., $c/\pi=0.8$), and (d) in deep chaotic regime (i.e., $c/\pi=1.4$). 
The solid black line is the exact solution $F_Q(t)$, the dashed line is the BMF approach $F_B(t)$, while the $F_{\rm HP}$ is the semiclassical HP Eq.~(\ref{HPOATjiexi}). 
We have set $A/\pi=0.4$ and $N=400$.}\label{newfig9}
 \end{figure} 

\begin{figure*}[t]
\includegraphics[width=0.75\linewidth]{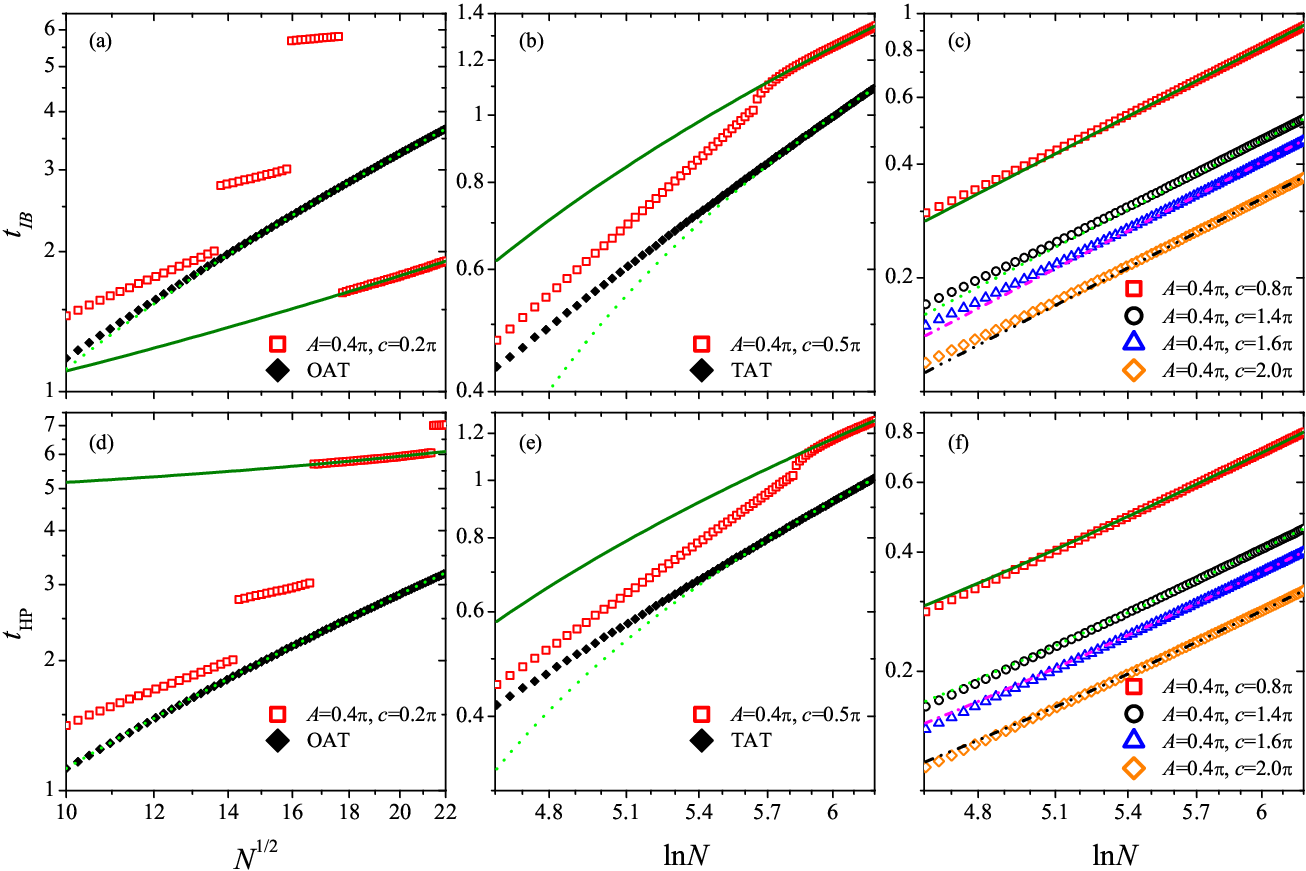}
\caption{Top [Bottom] row panels show the break time $t_{IB}$,
as defined in Eq.~(\ref{tIB}) [$t_{\rm HP}$,  Eq.~(\ref{tHP})], as a function of the total particle number $N$ for the QKR model.
The different panels corresponds to the initial state being 
(a,d) a classical stable point, (b,e) a classical saddle point, and (c,f) in the deep chaotic region. 
In all panels, symbols are numerical results while lines are fits, see text.
Values of Hamiltonian parameters are reported in each panel, while parameters of the parameters of the fitting lines are reoported in the text.
Panel (a,d) shows the results for the OAT model (black diamonds) and the QKR model in the stable region (red squares).
Panel (b,e) shows the results for the TAT model (black diamonds) and the QKR model at the saddle point (red squares).
Panel (c,f) shows the results of the QKR model in the chaotic regime. 
}\label{fig3}
 \end{figure*}

\begin{figure}[t]
\includegraphics[width=\linewidth]{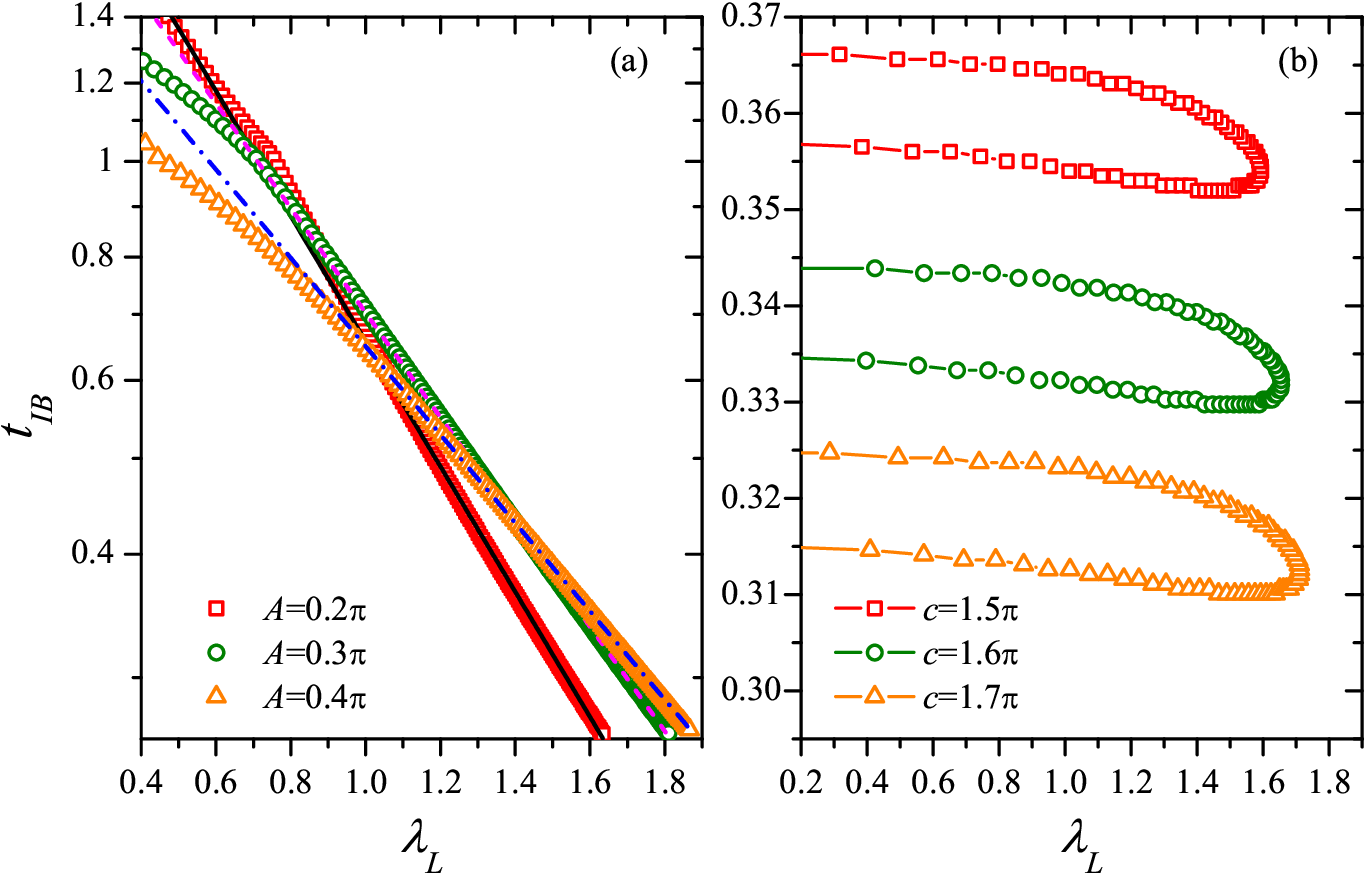}
\caption{Break time $t_{IB}$ as a function of the maximum Lyapunov exponent $\lambda_L$. 
Symbols are numerical results, lines are fits according to Eq.~(\ref{prefactorchaos}).
In panel (a), the parameter $A$ is fixed and the Lyapunov exponent only depends on $c$.
The fitting lines are $2.83\times{e}^{-1.46\lambda_L}$ (solid line), $2.38\times{e}^{-1.22\lambda_L}$ (dash line), and  $1.82\times{e}^{-1.03\lambda_L}$ (dash-dot line), respectively. 
In panel (b), the parameter $c$ is fixed and the maximum Lyapunov exponent is a double-valued function of $A$. The display of the lines between the scattered points is only for guiding eyes.  
The initial state is $s_x(0)=1$ and we set $N=300$.}
\label{fignew4}
 \end{figure}

In Fig. \ref{newfig9}, we compare the exact $F_Q(t)$ with both the BMF approximation $F_B(t)$ and the HP approximation $F_{\rm HP}(t)$ 
for the periodic-kicked system with the initial state $s_x(0)=1$. 
When the initial state is a classical stable point, the three QFI agree well for a long time [see Fig. \ref{newfig9}(a)]. 
When the initial state is an unstable saddle point [see Fig. \ref{newfig9}(b)] or in the classical chaotic region [see Figs. \ref{newfig9}(c) and \ref{newfig9}(d)]
the three results remain consistent only for a short time (a few kicks in the figure). 
While the $F_{\rm HP}(t)$ diverges after a transient time, the BMF follows more closely the exact results for longer times. 
It is thus interesting to introduce the criterion 
\be \label{tIB}
\frac{|F_Q(t_{IB})-F_B(t_{IB})|}{F_Q(t_{IB})} = g_{IB},
\ee 
to quantitatively define a break time $t_{IB}$ where the BMF approximation departs from the exact result
(by a relative amount $g_{IB}$).
This time is closely related to the instability degree of the initial state, which we will relate below to the maximum Lyapunov exponent. 
In practice, we set $g_{IB}=0.01$ in the following.
The main result of this paper is to study numerically the break time $t_{IB}$ in different chaotic regimes and determine characteristic behaviours.
We consider different regimes:
\begin{itemize}

\item[$\bullet$] If the initial condition is in the classical stable region, we find 
\be \label{tIBstable}
t_{IB}\simeq \alpha_{\rm st}(A,c) \times \sqrt{N}.  
\ee
In particular, we can demonstrate analytically that the $O(\sqrt{N})$ behaviour holds for the OAT dynamics.
To support the claim (\ref{tIBstable}) we show numerical results in Fig.~\ref{fig3}(a), obtained for the QKR model for $A/\pi=0.4$ and $c/\pi = 0.2$.  
There, we plot the break time $t_{IB}$ as a function of the number of particles $N$.
The dashed line in Fig.~\ref{fig3}(a) is $\sqrt{N}/4.7$ fitting the OAT results, while the solid line is $\sqrt{N}/15$ fitting the QKR results for large $N$, in agreement with Eq.~(\ref{tIBstable}).

\item[$\bullet$]  If the initial condition is at the saddle point, we find 
\begin{equation} \label{tIBsaddle}
t_{IB}\simeq \alpha_{\rm sp}(A,c) \times \ln{N}.   
\end{equation}
In particular, this behaviour holds for the TAT dynamics as well as in the QKR model at the saddle point [see Fig. \ref{fig3}(b)].
The dashed line in Fig.~\ref{fig3}(b) is $\ln N/2.02$ fitting the TAT results, while the solid line is $\ln N/2.2$ fitting the QKR results for large $N$, in agreement with Eq.~(\ref{tIBsaddle}).
Notice that, for the TAT model with initial conditions at the saddle point, we expect 
$t_{IB} \propto t_c$, where $t_c \sim (1/\lambda) \log N$ (for $N \gg 1$), as given in Eq.~(\ref{TATtc}), with $\lambda = \sqrt{A(c-A)}$.

\item[$\bullet$]  If the initial conditions are in the deep chaotic regime we find
\begin{equation} \label{tIBchaos}
t_{IB}\simeq \alpha_{\rm ch}(A,c) \times (\ln{N})^4,
\end{equation}
see Fig. \ref{fig3}(c).
The solid line in Fig.~\ref{fig3}(c) is the fit $(\ln N)^4/1590$, the dashed line is $(\ln N)^4/2810$, the dash-dotted line is $(\ln N)^4/3200$, and the dash-double-dotted line is $(\ln N)^4/4000$. 
In the chaotic regime, all results agree well with Eq.~(\ref{tIBchaos}).
Notice that we numerically do not see a smooth transition between the $O(\log N)$ behaviour at the saddle point and the $O((\log N)^4)$ behaviour in the deep chaotic regime, 
but rather a sharp transition.
Furthermore, we can connect the prefactor $\alpha_{\rm ch}(A,c)$ in Eq.~(\ref{tIBchaos}) and the asymptotically maximum Lyapunov exponent $\lambda_L(A,c)$ 
(depending on both parameters $c$ and $A$).
For strong chaos and large $N$, we find numerically
\begin{equation} \label{prefactorchaos}
\alpha(A,c)_{\rm ch} \simeq\eta(A){e}^{-\gamma(A)\lambda_L(A,c)},    
\end{equation}
where the coefficients $\gamma(A)$ and $\eta(A)$ depend only on the kick strength $A$.
A numerical investigation of the Eq.~(\ref{prefactorchaos}) is shown in Fig.~\ref{fignew4}. In Fig. \ref{fignew4}(a) we show the break time 
as a function of $\lambda_L(A,c)$ and 
for different values of kicked strength, $A/\pi=0.2$, $0.3$, and $0.4$.
Equation (\ref{prefactorchaos}) predicts that, 
while fixing $A$, the break time $t_{IB}$ depends on $c$ through an exponential function of the maximum Lyapunov exponent. 
This behaviour (lines) is well reproduced by the numerical calculations (symbols) for sufficiently large values of $\lambda_L(A,c)$, as shown in Fig. \ref{fignew4}(a).
Furthermore, when fixing the parameter $c$,
Fig. \ref{fig1}(a) shows that the maximum Lyapunov exponent in the chaotic region is a double-valued function of $A$.
Therefore, Eq.~(\ref{prefactorchaos}) also predicts that $t_{IB}$ is double-valued function of $A$, for fixed $c$.
This prediction is confirmed in Fig. \ref{fignew4}(b), where we plot $t_{IB}$ as a function of $\lambda_L(A,c)$ for different values of $c$.
\end{itemize}

In Ref.~\cite{LerosePRA2020b}, the time evolution of the HP approximation to the QFI, $F_{\rm HP}(t)$, 
has been related to the local (i.e., finite-time) maximum Lyapunov exponent $\lambda_t$, showing that $F_{\rm HP}(t) \sim e^{2\lambda_t t}$.
This relation holds in a transient time window, up to the Ehrenfest time.  
We notice that the HP approach of Ref.~\cite{LerosePRA2020b} agrees well with our results.
In Fig.~\ref{fig3} we plot the break time of the HP approximation, $t_{\rm HP}$, defined in analogy to $t_{IB}$ as
\be \label{tHP}
\frac{|F_Q(t_{\rm HP})-F_B(t_{\rm HP})|}{F_Q(t_{\rm HP})} = g_{\rm HP}.
\ee 
Panels (d)-(f) of Fig.~\ref{fig3} show $t_{\rm HP}$ as a function of $N$.
Panel (d) corresponds to the initial state in the
classical stable region.
Lines in Fig.~\ref{fig3}(d) are numerical fit $t_{\rm HP} = O(\sqrt{N})$:
the dashed line is $\sqrt{N}/5.8$ fitting the OAT results, while the solid line is $\sqrt{N}/13$ fitting the QKR results for large $N$.
Panel (e) corresponds to the saddle point, where $t_{\rm HP} = O(\log N)$: the dashed line is $\ln N/2.34$ fitting the TAT results, while the solid line is $\ln N/2.34$ fitting the QKR results for large $N$.
Panel (f) corresponds to the deep chaotic regime, where $t_{\rm HP} = O((\log N)^4)$:
the solid line is the fit $(\ln N)^4/2020$, the dashed line is $(\ln N)^4/3540$, the dash-dotted line is $(\ln N)^4/4070$, and the dash-double-dotted line is $(\ln N)^4/5100$.
The scaling with $N$ is thus analogous to Eqs.~(\ref{tIBstable}), (\ref{tIBsaddle}) and (\ref{tIBchaos}), respectively.

\begin{figure}[b!]
\includegraphics[width=\linewidth]{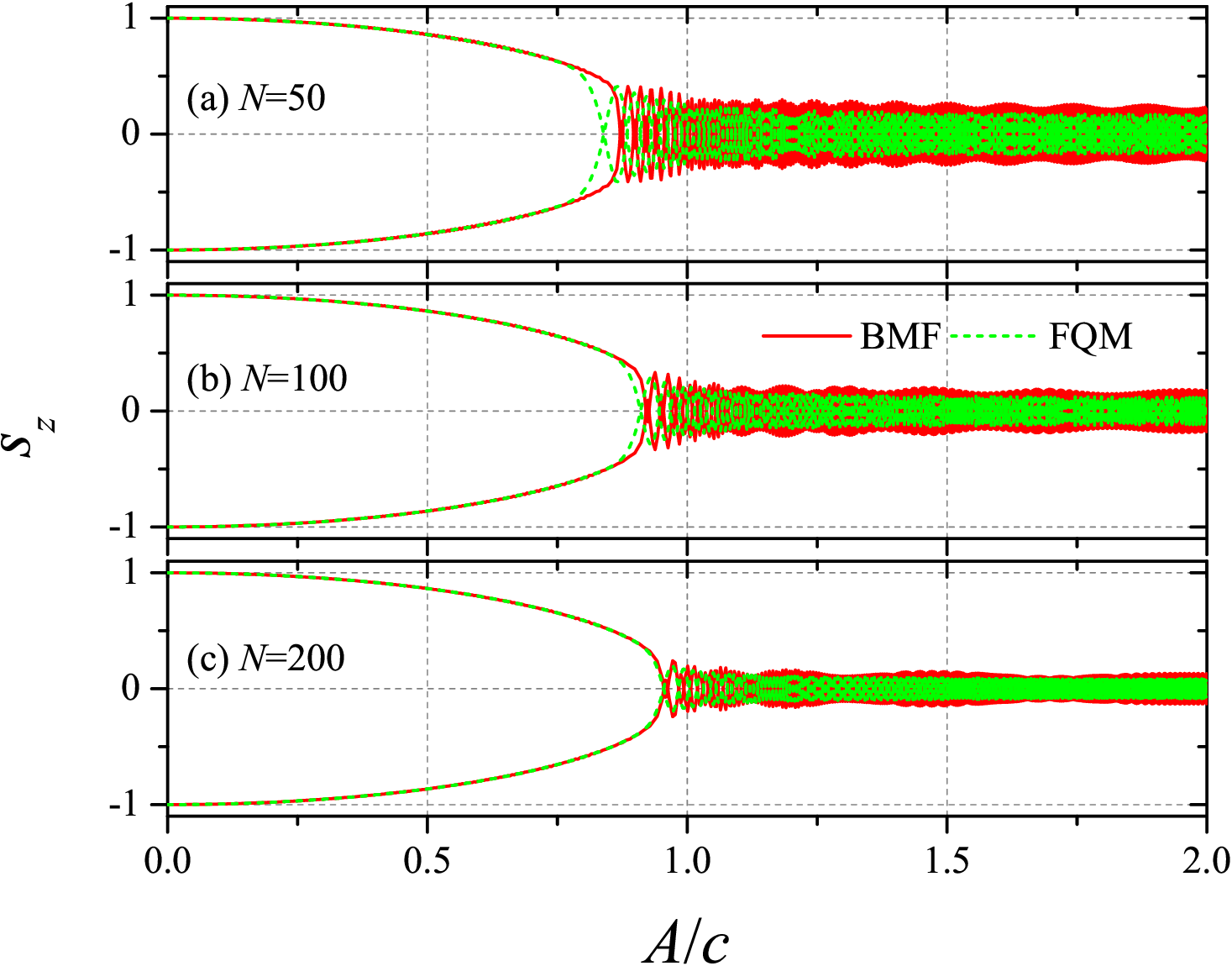}
\caption{Adiabatic dynamics comparison between full quantum model (FQM) and BMF model. To obtain the population imbalance probability $s_z(t)$, we adiabatically evolve the system from the initial ground states $s_z(0)=\pm 1$ according to both quantum Schr\"{o}dinger equation and BMF  equations with $A=v t$ ($v=10^{-3})$  and $c=1$.}\label{newfig6}
 \end{figure}   

 \section{Application in identifying a quantum phase transition}

 So far, we have shown how the BMF theory can describe the entanglement dynamics of quantum many-body kicked system up to transient times.
 Here, we discuss the application of the BMF approach to identify the critical point of a second-order quantum phase transition (QPT) that occurs in a bosonic Josephson junction.
 This QPT has been recently explored with a BEC with tuneable interparticle interaction~\cite{UlyanovPR1992, TrankwalderNATPHYS2016}
 (see Refs.~\cite{VidalPRA2004, MaPRA2009, HaukeNATPHY2016, GabbrielliSCIREP2018} for studies of entanglement in this system).
 The system is described by the Hamiltonian 
\begin{equation}\label{QPT}
\hat{H}=A\hat{J}_x-\frac{c}{N}\hat{J}_z^2,
\end{equation}
that is characterized by a QPT between 
a gapped symmetric state  (for $A/c>1$) to two degenerate asymmetric states  (i.e., $A/c<1$), with critical point (in the thermodynamic limit) at $A/c=1$. 
Instead of direct diagonalization, here we use adiabatic dynamical method to describe and identify the transition between the two regimes in a system with finite number of particles. Initially, we assume that the coupling between the two modes is very weak, and the eigenstates of $\hat{J}_z$, i.e., $|\mu=\pm{N}/{2}\rangle$ (or $s_z(0)=\pm 1$) are the ground state of the system, which are double degenerate, where all atoms occupy in one mode. 
Then we adiabatically evolve the system by slowly increasing the coupling strength between the two modes as $A=v t$. If the coupling strength changes slowly enough, i.e., $v\rightarrow 0$, the quantum adiabatic theorem ensures that the system is always in the instantaneous ground state. When the coupling exceeds a certain pseudo-phase transition critical point, the ground state of the system will change from mainly occupying a single mode  to having two modes equally occupied. This adiabatic dynamics shown in Fig. \ref{newfig6} can be used to illustrate the process of phase transition. Obviously, increasing the number of particles, the consistency of the results between the full quantum model and the BMF model calculations becomes better and better. The oscillations of the order parameter $s_z$ for the phase transition after passing the pseudo-critical point will be suppressed with the decrease of adiabatic evolution velocity $v$.

\begin{figure}[t]
\includegraphics[width=\linewidth]{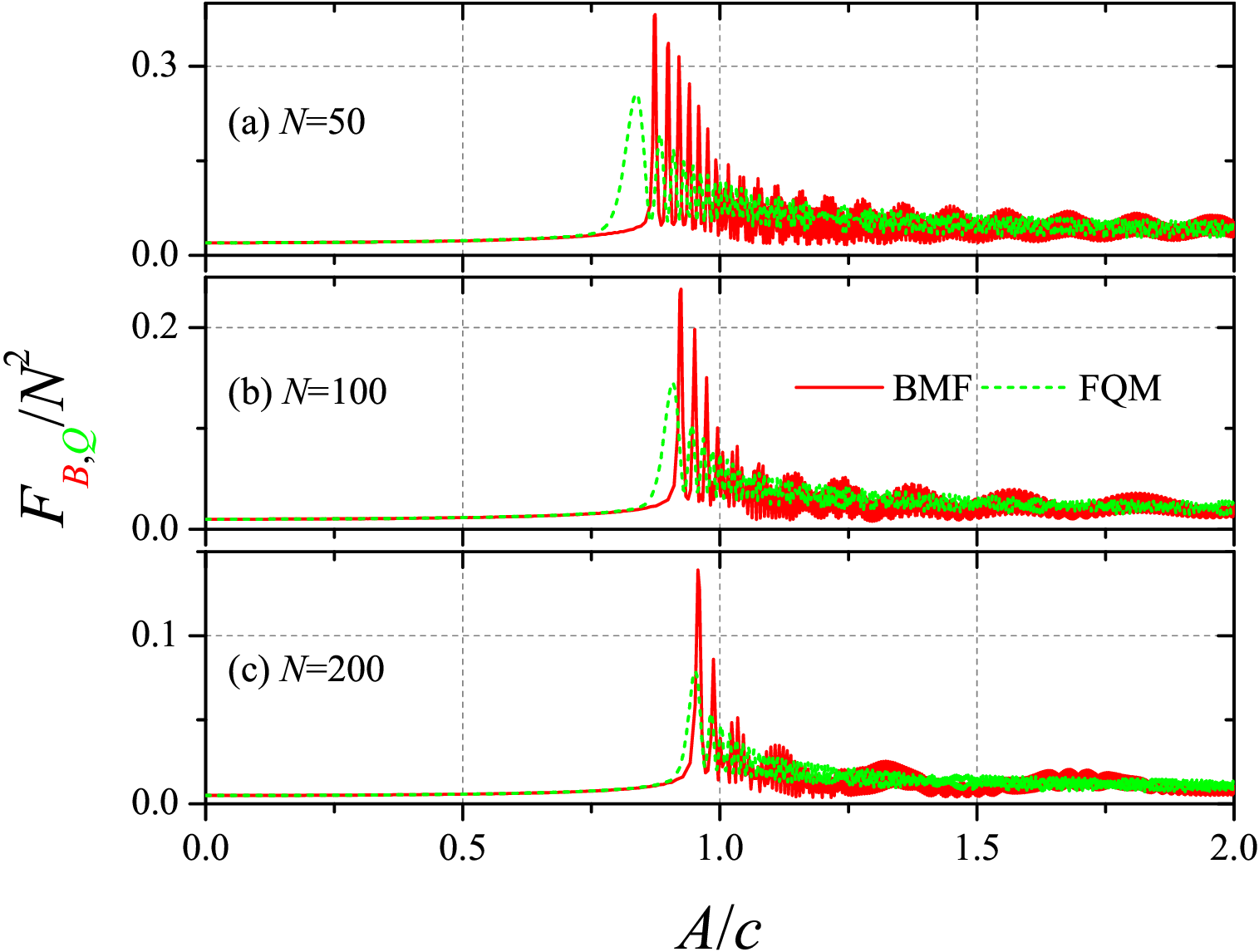}
\caption{QFI comparison between full quantum model (FQM) and BMF model. To obtain $F_{B,Q}$, we adiabatically evolve the system from the initial ground state $s_z(0)=1$ according to both quantum Schr\"{o}dinger equation and BMF equations with $A=v t$ ($v=10^{-3})$ and $c=1$.}\label{newfig5}
 \end{figure}  

The corresponding QFI of the instantaneous ground state during adiabatic evolution is also calculated and the results are shown in Fig. \ref{newfig5}. We find that the QFI has a maximum value at the pseudo-critical point of phase transition. This can be used as an effective signal to recognize the phase transition of finite size systems. It is seen that the BMF approximation becomes more precise when cranking up the number $N$. This means that for quantum many-body system with large but finite number of particles, the simple BMF theory can be used to accurately predict the position of phase transition without solving the complex quantum many-body model.

\section{Conclusion}

To conclude, we have investigated the entanglement dynamics of a quantum kicked rotor model by using a beyond mean-field approach. 
We have found numerically characteristic scaling laws for the break time $t_{IB}$ when the quantum Fisher information calculated within the BMF approach 
departs from the exact value. 
The scaling depends on the properties of the initial conditions: 
if the initial conditions correspond to a classical regular regime, $t_{IB} = O(\sqrt{N})$; 
 if the initial conditions correspond to a classical saddle point, $t_{IB} = O(\log N)$; otherwise, 
if the initial conditions correspond to a classical chaotic regime, $t_{IB} = O((\log N)^4)$.
{We thus conclude that the entanglement dynamics of unstable saddle point is the most irregular, in the sense that the moment when the quantum effects cannot be captured by the BMF approach occurs earlier than in both regular and chaotic evolution.} 
Furthermore, in the chaotic regime, we have shown a relation between $t_{IB}$ and the maximum Lyapunov exponent,  $t_{IB} \sim e^{-\gamma\lambda_L}\times(\log N)^4$, see Eqs.~(\ref{tIBchaos}) and ({\ref{prefactorchaos}}).
In particular, we have found that the QFI has a maximum value at the pseudo-critical point of phase transition in a bosonic Josephson junction, which suggests an application in accurately identifying the critical point of quantum phase transition in finite-size systems.
While our study is restricted for pure states and it is expected to hold for vanishing decoherence, a study of mixed states will be discussed in a future work.

 \section*{Acknowledgement}

We acknowledge discussions with Alessio Lerose and 
Silvia Pappalardi.
This work is supported by the National Natural Science Foundation of China (Grant No. 11974273) and the Natural Science Fundamental Research Program of Shaanxi Province of China (Grant No. 2019JM-004).


\begin{thebibliography}{99}
\bibitem{Wigner1951} 
E. P. Wigner, 
On the statistical distribution of the widths and spacings of nuclear resonance levels, 
Math. Proc. Cambr. Philos. Soc. {\bf 47}, 790 (1951).

\bibitem{Einstein1917} 
A. Einstein, Zum Quantensatz von Sommerfeld und Epstein, 
Verh. Dtsch. Phys. Ges. {\bf 19}, 82 (1917). 
Reprinted in The Collected Papers of Albert Einstein, A. Engel translator (Princeton University Press, Princeton, 1997).

\bibitem{Bychek2019}
A. A. Bychek, P. S. Muraev, and A. R. Kolovsky, 
Probing quantum chaos in many-body quantum systems by the induced dissipation, Phys. Rev. A {\bf 100}, 013610 (2019).

\bibitem{ChaudhuryNATURE2009}
S. Chaudhury, A. Smith, B. Anderson, S. Ghose, and P. S. Jessen. 
Quantum signatures of chaos in a kicked top. 
Nature {\bf 461}, 768-771 (2009).








\bibitem{Stockmann1990}
H.-J. St\"{o}ckmann and J. Stein, 
``Quantum" Chaos in Billiards Studied by Microwave Absorption, 
Phys. Rev. Lett. {\bf 64}, 2215 (1990).

\bibitem{Graf1992}
H.-D. Gr\"{a}f, H. L. Harney, H. Lengeler, C. H. Lewenkopf, C. Rangacharyulu, A. Richter, P. Schardt, and H. A. Weidenm\"{u}ller, 
Distribution of Eigenmodes in a Superconducting Stadium Billiard with Chaotic Dynamics, 
Phys. Rev. Lett. {\bf 69}, 1296 (1992).

\bibitem{Milner2001}
V. Milner, J. L. Hanssen, W. C. Campbell, and M. G. Raizen, 
Optical Billiards for Atoms, 
Phys. Rev. Lett. {\bf 86}, 1514 (2001).

\bibitem{Friedman2001}
N. Friedman, A. Kaplan, D. Carasso, and N. Davidson, 
Observation of Chaotic and Regular Dynamics in Atom-Optics Billiards, 
Phys. Rev. Lett. {\bf 86}, 1518 (2001).

\bibitem{Zhang2008} 
Q. Zhang, P. Hanggi, and J. B. Gong, 
Nonlinear Landau-Zener processes in a periodic driving field, 
New J. Phys. {\bf 10}, 073008 (2008).

\bibitem{Lepers2008} 
M. Lepers, V. Zehnle, and J. C. Garreau, 
Tracking Quasiclassical Chaos in Ultracold Boson Gases, 
Phys. Rev. Lett. {\bf 101}, 144103 (2008).

\bibitem{Liu2006} 
J. Liu, C. Zhang, M. G. Raizen, and Q. Niu, 
Transition to instability in a periodically kicked Bose-Einstein condensate on a ring, 
Phys. Rev. A {\bf 73}, 013601 (2006).

\bibitem{Martin2008} 
J. Martin, B. Georgeot, and D. L. Shepelyansky, 
Time Reversal of Bose-Einstein Condensates, 
Phys. Rev. Lett. {\bf 101}, 074102 (2008).

\bibitem{Kidd2019}
R. A. Kidd, M. K. Olsen, and J. F. Corney, 
Quantum chaos in a Bose-Hubbard dimer with modulated tunneling, 
Phys. Rev. A {\bf 100}, 013625 (2019).

\bibitem{Coullet2002} 
P. Coullet and N. Vandenberghe, Chaotic dynamics of a Bose-Einstein condensate in a double-well trap, 
J. Phys. B: At. Mol. Opt. Phys. {\bf 35}, 1593 (2002).

\bibitem{Cheng2010}
J. Cheng, 
Chaotic dynamics in a periodically driven spin-1 condensate, 
Phys. Rev. A {\bf 81}, 023619 (2010).

\bibitem{Kronjager2008} 
J. Kronj\"{a}ger, K. Sengstock, and K. Bongs, 
Chaotic dynamics in spinor Bose-Einstein condensates, New J. Phys. {\bf 10}, 045028 (2008).

\bibitem{Cheng2009} 
J. Cheng, 
Regular and irregular spin-mixing dynamics in coupled spin-1 atomic and molecular Bose-Einstein condensates, 
Phys. Rev. A {\bf 80}, 023608 (2009).

\bibitem{Haake1987}
F. Haake, M. Kuś, and R. Scharf, 
Classical and quantum chaos for a kicked top
Z. Phys. B Condens. Matter {\bf 65}, 381 (1987).
 
\bibitem{HaakeBOOK}
F. Haake, 
{\it Quantum Signatures of Chaos} (Springer, Berlin, 2010).

\bibitem{Moore1995} 
F. L. Moore, J. C. Robinson, C. F. Bharucha, B. Sundaram, and M. G. Raizen, 
Atom Optics Realization of the Quantum $\delta$-Kicked Rotor, 
Phys. Rev. Lett. {\bf 75}, 4598 (1995).

\bibitem{Fallani2004} 
L. Fallani, L. De Sarlo, J. E. Lye, M. Modugno, R. Saers, C. Fort, and M. Inguscio, 
Observation of Dynamical Instability for a Bose-Einstein Condensate in a Moving 1D Optical Lattice, 
Phys. Rev. Lett. {\bf 93}, 140406 (2004).

\bibitem{Christiani2004} 
M. Christiani, O. Morsch, N. Malossi, M. Jona-Lasinio, M. Anderlini, E. Courtade, and E. Arimondo, 
Instabilities of a Bose-Einstein condensate in a periodic potential: an experimental investigation, 
Opt. Express {\bf 12}, 4 (2004).

\bibitem{Duffy2004} 
G. J. Duffy, S. Parkins, T. Muller, M. Sadgrove, R. Leonhardt, and A. C. Wilson, 
Experimental investigation of early-time diffusion in the quantum kicked rotor using a Bose-Einstein condensate, 
Phys. Rev. E {\bf 70}, 056206 (2004).

\bibitem{Creffield2006} 
C. E. Creffield, G. Hur, and T. S. Monteiro, 
Localization-Delocalization Transition in a System of Quantum Kicked Rotors, 
Phys. Rev. Lett. {\bf 96}, 024103 (2006).

\bibitem{Chirikov1979} 
B. V. Chirikov, 
A universal instability of many-dimensional oscillator systems, 
Phys. Rep. {\bf 52}, 263 (1979).

\bibitem{Stockmann1999} 
H. J. Stockmann, 
{\it Quantum Chaos: An Introduction} (Cambridge University Press, Cambridge, 1999).

\bibitem{Lewis-Swan2019} 
R. Lewis-Swan, A. Safavi-Naini, J. Bollinger, and A. Rey, 
Unifying scrambling, thermalization and entanglement through measurement of fidelity out-of-time-order correlators in the Dicke model, 
Nat. Comm. {\bf 10}, 1581 (2019).

\bibitem{GietkaPRB2019}
K. Gietka, J. Chwedeńczuk, T. Wasak, and F. Piazza,
Multipartite entanglement dynamics in a regular-to-ergodic transition: Quantum Fisher information approach,
Phys. Rev. B {\bf 99}, 064303 (2019).

\bibitem{LerosePRA2020b}
A. Lerose and S. Pappalardi, 
Bridging entanglement dynamics and chaos in semiclassical systems, 
Phys. Rev. A {\bf 102}, 032404 (2020).

\bibitem{PezzePRL2009}
L. Pezz\`{e} and A. Smerzi, 
Entanglement, Nonlinear Dynamics, and the Heisenberg Limit, 
Phys. Rev. Lett. {\bf 102}, 100401 (2009).

\bibitem{note_multipartite}
The notion of multipartite  entanglement  is related to that of $k$-producibility. A pure state is $k$-partite entangled if it is not $(k-1)$-producible, namely if it can be written as a product $\otimes_j \ket{\psi_j}$ 
containing at least one state $\ket{\psi_j}$ of $k-1$ particles that does not factorize. 
The definition is then extended to mixed states by convexity, see~\cite{TothPR2009, HyllusPRA2012, TothPRA2012}.

\bibitem{TothPR2009}
O G\"uhne and G T\'oth,
Entanglement detection,
Phys. Rep. {\bf 474}, 1 (2009).

\bibitem{HyllusPRA2012}
P. Hyllus, W. Laskowski, R. Krischek, C. Schwemmer, W. Wieczorek, H. Weinfurter, L. Pezz\`{e}, and A. Smerzi, 
Fisher information and multiparticle entanglement, 
Phys. Rev. A {\bf 85}, 022321 (2012).

\bibitem{TothPRA2012}
G. T\'oth, 
Multipartite entanglement and high-precision metrology, 
Phys. Rev. A {\bf 85}, 022322 (2012).

\bibitem{TothJPA2014}
G. T\'oth and I. Apellaniz, 
Quantum metrology from a quantum information science perspective, 
J. Phys. A {\bf 47}, 424006 (2014).

\bibitem{PezzePNAS2016}
L. Pezz\`e, Y. Li, W. Li, and A. Smerzi, 
Witnessing entanglement without entanglement witness operators, 
Proc. Natl. Acad. Sci. U.S.A. {\bf 113}, 11459 (2016).

\bibitem{PezzeRMP2018}
L. Pezz\`{e}, A. Smerzi, M. K.  Oberthaler, R. Schmied, and P. Treutlein, 
Quantum metrology with nonclassical states of atomic ensembles, 
Rev. Mod. Phys. {\bf 90}, 035005 (2018).

\bibitem{HelstromBOOK}
C. W. Helstrom, {\it  Quantum Detection and Estimation Theory} (Academic Press, New York, 1976).

\bibitem{CavesPRL1994}
S. L. Braunstein, and  C. M. Caves, 
Statistical distance and the geometry of quantum states, 
Phys. Rev. Lett. {\bf 72}, 3439 (1994).

\bibitem{MaPRA2009}
J. Ma and X. Wang, 
Fisher information and spin squeezing in the Lipkin-Meshkov-Glick model,
Phys. Rev. A {\bf 80} 012318 (2009).

\bibitem{HaukeNATPHY2016}
P. Hauke, M. Heyl, L. Tagliacozzo, and P. Zoller, 
Measuring multipartite entanglement through dynamic susceptibilities. 
Nat. Phys. {\bf 12}, 778 (2016).

\bibitem{PezzePRL2017}
L. Pezz\`e, M. Gabbrielli, L. Lepori and A. Smerzi, 
Multipartite entanglement in topological quantum phases, Phys.Rev.Lett. {\bf 119} 250401 (2017)

\bibitem{PappalardiJSM2017}
S. Pappalardi, S. Russomanno, A. Silva, and R. Fazio,  Multipartite entanglement after a quantum quench, 
J. Stat. Mech. {\bf 2017}, 053104 (2017).

\bibitem{GabbrielliSCIREP2018}
M. Gabbrielli, A. Smerzi and L. Pezz\`e, 
Multipartite Entanglement at Finite Temperature,
Sci. Rep. {\bf 8}, 15663 (2018).

\bibitem{GabbrielliNJP2019}
M. Gabbrielli, L. Lepori and L. Pezz\`e, 
Multipartite-entanglement tomography of a quantum simulator
New J. Phys. {\bf 21}, 033039 (2019).

\bibitem{FrerotNATCOMM2019}
I. Fr\'erot and T. Roscilde, 
Reconstructing the quantum critical fan of strongly correlated systems using quantum correlations,
Nat. Comm {\bf 10}, 577 (2019). 

\bibitem{SorelliPRA2019}
G. Sorelli, M. Gessner, A. Smerzi, and L. Pezz\`{e}, 
Fast and optimal generation of entanglement in bosonic Josephson junctions, 
Phys. Rev. A {\bf 99}, 022329 (2019).

\bibitem{AnglinPRA2001b}
J. R. Anglin and A. Vardi, 
Dynamics of a two-mode Bose-Einstein condensate beyond mean-field theory, 
Phys. Rev. A {\bf 64}, 013605 (2001).

\bibitem{Fiderer2018}
L. J. Fiderer and D. Braun, 
Quantum metrology with quantum-chaotic sensors, 
Nat. Commun. {\bf 9}, 1351 (2018).


\bibitem{Javanainen1986}
J. Javanainen, 
Oscillatory exchange of atoms between traps containing Bose condensates, 
Phys. Rev. Lett. {\bf 57}, 3164 (1986).

\bibitem{Milburn1997} 
G. J. Milburn, J. Corney, E. M. Wright, and D. F. Walls, 
Quantum dynamics of an atomic Bose-Einstein condensate in a double-well potential, 
Phys. Rev. A {\bf 55}, 4318 (1997).

\bibitem{Zapata1998} 
I. Zapata, F. Sols, and A. J. Leggett, Josephson effect between trapped Bose-Einstein condensates, 
Phys. Rev. A {\bf 57}, R28 (1998).

\bibitem{Smerzi1997} 
A. Smerzi, S. Fantoni, S. Giovanazzi, and S. R. Shenoy, 
Quantum coherent atomic tunneling between two trapped Bose-Einstein condensates, 
Phys. Rev. Lett. {\bf 79}, 4950 (1997).

\bibitem{JavanainenPRA1999}
 J. Javanainen and M. Yu. Ivanov, 
 Splitting a trap containing a Bose-Einstein condensate: Atom number fluctuations,
 Phys. Rev. A {\bf 60}, 2351 (1999).

\bibitem{AlbiezPRL2005}
M. Albiez, R. Gati, J. F\"olling, S. Hunsmann, M. Cristiani, and M. K. Oberthaler, 
Direct observation of tunneling and nonlinear self-trapping in a single bosonic Josephson junction, 
Phys. Rev. Lett. {\bf 95}, 010402 (2005).

\bibitem{SchummNATPHYS2005}
T. Schumm, S. Hofferberth, L. M. Andersson, S. Wildermuth, S. Groth, I. Bar-Joseph, J. Schmiedmayer, and P. Kr\"uger, 
Matter-wave interferometry in a double well on an atom chip, 
Nat. Phys. {\bf 1}, 57 (2005).

\bibitem{Pezze2005} 
L. Pezz\`{e}, L. A. Collins, A. Smerzi, G. P. Berman, and A. R. Bishop, 
Sub-shot-noise phase sensitivity with a Bose-Einstein condensate Mach-Zehnder interferometer, 
Phys. Rev. A {\bf 72}, 043612 (2005).

\bibitem{Ananikian2006} 
D. Ananikian and T. Bergeman, 
Gross-Pitaevskii equation for Bose particles in a double-well potential: Two-mode models and beyond, 
Phys. Rev. A {\bf 73}, 013604 (2006).

\bibitem{Gati2007} 
R. Gati and M. K. Oberthaler, 
A bosonic Josephson junction, 
J. Phys. B: At., Mol. Opt. Phys. {\bf 40}, R61 (2007).

\bibitem{TrankwalderNATPHYS2016}
A. Trenkwalder, et al., 
Quantum phase transitions with parity-symmetry breaking and hysteresis, 
Nat. Phys. {\bf 12}, 826 (2016).

\bibitem{SpagnolliPRL2017}
G. Spagnolli, et al., 
Crossing over from attractive to repulsive interactions in a tunneling bosonic josephson junction, 
Phys. Rev. Lett. {\bf 118}, 230403 (2017).

\bibitem{Cirac1998} 
J. I. Cirac, M. Lewenstein, K. M\o lmer, and P. Zoller, 
Quantum superposition states of Bose-Einstein condensates, 
Phys. Rev. A {\bf 57}, 1208 (1998).

\bibitem{Steel1998} 
M. J. Steel and M. J. Collett, 
Quantum state of two trapped Bose-Einstein condensates with a Josephson coupling, 
Phys. Rev. A {\bf 57}, 2920 (1998).

\bibitem{ZiboldPRL2010}
T. Zibold, E. Nicklas, C. Gross, and M. K. Oberthaler, 
Classical bifurcation at the transition from Rabi to Josephson dynamics. 
Phys. Rev. Lett. {\bf 105}, 204101 (2010).

\bibitem{Gross2010} 
C. Gross, T. Zibold, E. Nicklas, J. Estève, and M. K. Oberthaler, 
Nonlinear atom interferometer surpasses classical precision limit, 
Nature {\bf 464}, 1165 (2010).

\bibitem{ArecchiPRA1972}
F. T. Arecchi, E. Courtens, R. Gilmore, and H. Thomas, 
Atomic Coherent States in Quantum Optics, 
Phys. Rev. A {\bf 6}, 2211 (1972).


\bibitem{FerriniPRA2011}
G. Ferrini, D. Spehner, A. Minguzzi, and F. W. J. Hekking, 
Effect of phase noise on quantum correlations in Bose-Josephson junctions,
Phys. Rev. A {\bf 84}, 043628 (2011).

\bibitem{Kitagawa1993} 
M. Kitagawa and M. Ueda, 
Squeezed spin states, 
Phys. Rev. A {\bf 47}, 5138 (1993).

\bibitem{SorensenNATURE2001} 
A. S. Sørensen, L.-M. Duan, J. I. Cirac, and P. Zoller, 
Many-particle entanglement with Bose-Einstein condensates, 
Nature {\bf 409}, 63 (2001).

\bibitem{Riedel2010} 
M. F. Riedel, P. B\"{o}hi, Y. Li, T. W. H\"{a}nsch, A. Sinatra, and P. Treutlein, 
Atom-chip-based generation of entanglement for quantum metrology, 
Nature {\bf 464}, 1170 (2010).

\bibitem{Schmied2016} 
R. Schmied, J.-D. Bancal, B. Allard, M. Fadel, V. Scarani, P. Treutlein, and N. Sangouard, 
Bell correlations in a Bose-Einstein condensate, 
Science {\bf 352}, 441 (2016).

\bibitem{Takeuchi2005} 
M. Takeuchi, S. Ichihara, T. Takano, M. Kumakura, T. Yabuzaki, and Y. Takahashi, 
Spin Squeezing Via One-Axis Twisting with Coherent Light, 
Phys. Rev. Lett. {\bf 94}, 023003 (2005).

\bibitem{Leroux2010} 
I. D. Leroux, M. H. Schleier-Smith, and V. Vuleti\'{c}, 
Implementation of Cavity Squeezing of a Collective Atomic Spin, 
Phys. Rev. Lett. {\bf 104}, 073602 (2010).

\bibitem{Schleier-Smith2010} 
M. H. Schleier-Smith, I. D. Leroux, and V. Vuleti\'{c}, 
Squeezing the collective spin of a dilute atomic ensemble by cavity feedback, 
Phys. Rev. A {\bf 81}, 021804 (2010).

\bibitem{BohnetNATPHOT2014}
J. G. Bohnet, K. C. Cox, M. A. Norcia, J. M. Weiner, Z. Chen, and J. K. Thompson, 
Reduced spin measurement back-action for a phase sensitivity ten times beyond the standard quantum limit, 
Nat. Photonics {\bf 8}, 731 (2014).

\bibitem{HostenNATURE2016}
O. Hosten, N. J. Engelsen, R. Krishnakumar, and M. A. Kasevich, 
Measurement noise 100 times lower than the quantum projection limit using entangled atoms, 
Nature {\bf 529}, 505 (2016).

\bibitem{MicheliPRA2003}
A. Micheli, D. Jaksch, J. I. Cirac, and P. Zoller, Many-particle entanglement in two-component Bose-Einstein condensates, 
Phys. Rev. A {\bf 67}, 013607 (2003).

\bibitem{StrobelSCIENCE2014} 
H. Strobel, W. Muessel, D. Linnemann, T. Zibold, D. B. Hume, L. Pezz\`{e}, A. Smerzi, and M. K. Oberthaler, 
Fisher information and entanglement of non-Gaussian spin states, 
Science {\bf 345}, 424 (2014).

\bibitem{Muessel2015} 
W. Muessel, H. Strobel, D. Linnemann, T. Zibold, B. Juli\'{a}-Diaz, and M. K. Oberthaler, 
Twist-and-turn spin squeezing in Bose-Einstein condensates, 
Phys. Rev. A {\bf 92}, 023603 (2015).

\bibitem{GarrtnerPRL2018}
M. G\"arrtner, P. Hauke, and A. M. Rey, 
Relating Out-of-Time-Order Correlations to Entanglement via Multiple-Quantum Coherences, 
Phys. Rev. Lett. {\bf 120}, 040402 (2018).

\bibitem{GirolamiENTROPY2017}
D. Girolami and B. Yadin, 
Witnessing Multipartite Entanglement by Detecting Asymmetry, 
Entropy {\bf 19}, 124 (2017).


\bibitem{Julia-DiazPRA2012a}
B. Juli\`a-D\'iaz, E. Torrontegui, J. Martorell, J. G. Muga, and A. Polls, 
Fast generation of spin-squeezed states in bosonic Josephson junctions, 
Phys. Rev. A {\bf 86}, 063623 (2012).

\bibitem{Julia-DiazPRA2012b}
B. Juli\`a-D\'iaz, T. Zibold, M.K. Oberthaler, M. Mele-Messeguer, J. Martorell, and A. Polls, 
Dynamic generation of spin- squeezed states in bosonic Josephson junctions, 
Phys. Rev. A {\bf 86}, 023615 (2012).

 \bibitem{ShchesnovichPRA2008}
V. S. Shchesnovich and M. Trippenbach, 
Fock-space WKB method for the boson Josephson model describing a Bose-Einstein condensate trapped in a double-well potential, 
Phys. Rev. A {\bf 78}, 023611 (2008).

\bibitem{LerosePRL2018}
A. Lerose, J. Marino, B. Zunkovic, A. Gambassi, and A. Silva,
Chaotic Dynamical Ferromagnetic Phase Induced by Nonequilibrium Quantum Fluctuations,
Phys. Rev. Lett. {\bf 120}, 130603 (2018);
A. Lerose, B. Zunkovic, J. Marino, A. Gambassi, and A. Silva,
Impact of nonequilibrium fluctuations on prethermal dynamical phase transitions in long-range interacting spin chains, 
Phys. Rev. B {\bf 99}, 045128 (2019).


\bibitem{note1}
The QFI is given by the maximum between Eq.~(\ref{FQjiexi}) and 
\begin{equation}\label{FQjiexi1}
\frac{F_Q(t)}{N}=N\left(1-\cos^{2N-2}\frac{ct}{N}\right)-(N-1)\frac{\alpha_Q(t)}{2}.
\end{equation}
Equation (\ref{FQjiexi1}) is larger than Eq.~(\ref{FQjiexi}) for $ct \simeq N\pi/2 -2\sqrt{N}$ and even values of $N$.

\bibitem{LerosePRR2020a}
A. Lerose and S. Pappalardi, 
Origin of the slow growth of entanglement entropy in long-range interacting spin systems, 
Phys. Rev. Research {\bf 2}, 012041(R) (2020).

\bibitem{LerosePRIVATE}
A. Lerose and S. Pappalardi, private communication.
S. Pappalardi, {\it Entanglement dynamics and chaos in long-range quantum systems} PhD thesis (2020). Available at https://iris.sissa.it.

\bibitem{LawPRA2001}
C. K. Law, H. T. Ng, and P. T. Leung, 
Coherent control of spin squeezing, 
Phys. Rev. A {\bf 63}, 055601 (2001). 

\bibitem{Liu2005}
J. Liu, W. Wang, C. Zhang, Q. Niu, and B. Li, 
Fidelity for the quantum evolution of a Bose-Einstein condensate, 
Phys. Rev. A {\bf 72}, 063623 (2005).

\bibitem{UlyanovPR1992}
V. V. Ulyanov and O. B.  Zaslavskii, 
New methods in the theory of quantum spin systems, 
Phys. Rep. {\bf 216}, 179 (1992).

\bibitem{VidalPRA2004}
J. Vidal, G. Palacios and R. Mosseri, 
Entanglement in a second-order quantum phase transition,
Phys. Rev. A {\bf 69}, 022107 (2004).

\end{thebibliography}
\end{document}